%% file: main_v0.5.tex
\documentclass[
 aps, reprint,
 superscriptaddress,
 longbibliography, nopreprintnumbers, noeprint,
 amsmath, amssymb, floatfix
]{revtex4-1}

\usepackage{multirow}
\usepackage{graphicx}
\usepackage{dcolumn}
\usepackage{bm}
\usepackage{dsfont}
\usepackage{hyperref}
\usepackage{physics}
\usepackage[usenames,dvipsnames]{xcolor}
\usepackage{bbold}
\usepackage[normalem]{ulem} 
\usepackage[inline]{enumitem}
\usepackage{qcircuit}
\usepackage{siunitx}
\usepackage{tikz}
\usepackage{amssymb} 
\usepackage[caption=false]{subfig}
\usepackage{mathrsfs}
\DeclareMathAlphabet{\mathpzc}{OT1}{pzc}{m}{it}

\newcommand{\sx}{\hat{\sigma}_x}
\newcommand{\sy}{\hat{\sigma}_y}
\newcommand{\sz}{\hat{\sigma}_z}
\newcommand{\xax}{\boldsymbol{\hat{x}}}
\newcommand{\yax}{\boldsymbol{\hat{y}}}

\newcommand{\h}{\hat{\mathcal{H}}}
\newcommand{\lhat}{\hat{L}}
\newcommand{\thetafull}{\theta_\text{full}}
\newcommand{\thetagate}{\theta_\text{gate}}

\newcommand{\lin}{\mathcal{L}}

\newcommand{\figref}[1]{Fig. \ref{#1}}
\newcommand{\secref}[1]{Section \ref{#1}}
\newcommand{\appref}[1]{Appendix \ref{#1}}
\newcommand{\tabref}[1]{Table \ref{#1}}
\renewcommand{\eqref}[1]{Eq. \ref{#1}}

\begin{document}

\title{Modelling non-Markovian noise in driven superconducting qubits}

\author{Abhishek Agarwal}
\email{abhishek.agarwal@npl.co.uk}
\author{Lachlan P. Lindoy}
\author{Deep Lall}
\author{Fran\c{c}ois Jamet}
\author{Ivan Rungger}
\email{ivan.rungger@npl.co.uk}
\affiliation{National  Physical  Laboratory,  Teddington,  TW11  0LW,  United  Kingdom}

\begin{abstract}
  Non-Markovian noise can be a significant source of errors in superconducting qubits. We develop gate sequences utilising mirrored pseudoidentities that allow us to characterise and model the effects of non-Markovian noise on both idle and driven qubits. We compare three approaches to modelling the observed noise: (i) a Markovian noise model, (ii)  a model including interactions with a two-level system (TLS), (iii) a model utilising the post Markovian master equation (PMME), which we show to be equivalent to the qubit-TLS model in certain regimes. When running our noise characterisation circuits on a superconducting qubit device we find that purely Markovian noise models cannot reproduce the experimental data. Our model based on a qubit-TLS interaction, on the other hand, is able to closely capture the observed experimental behaviour for both idle and driven qubits. We investigate the stability of the noise properties of the hardware over time, and find that the parameter governing the qubit-TLS interaction strength fluctuates 
 significantly even over short time-scales of a few minutes. Finally, we evaluate the changes in the noise parameters when increasing the qubit drive pulse amplitude. We find that although the hardware noise parameters fluctuate significantly over different days, their drive pulse induced relative variation is rather well defined within computed uncertainties: both the phase error and the qubit-TLS interaction strength change significantly with the pulse strength, with the phase error changing quadratically with the amplitude of the applied pulse. Since our noise model can closely describe the behaviour of idle and driven qubits, it is ideally suited to be used in the development of quantum error mitigation and correction methods.
  
\end{abstract}

\maketitle
\section{Introduction}
Superconducting qubits are among the leading technologies in the development of a large-scale universal quantum computer \cite{kjaergaard2020superconducting,krantzQuantumEngineerGuide2019}. To achieve the goal of an error-corrected quantum computer, the number of available qubits needs to be increased significantly, while also reducing the effects of noise in the device. Understanding the physical sources of noise in the device, and the effect they have on qubits, is vital to the development of new devices with reduced noise levels, and of new methods that mitigate and correct the effects of noise in the device. 

Various types of noise affect superconducting qubit devices. Noise caused due to interactions with the environment, such as charge noise, photon number fluctuations, and quasiparticles, can lead to energy relaxation and pure dephasing in the qubit \cite{oliver2013materials,krantzQuantumEngineerGuide2019, burnettDecoherenceBenchmarkingSuperconducting2019}. Detailed characterization of a qubit's interaction with its environment is a challenging task that requires accurate noise spectroscopy experiments \cite{burnettEvidenceInteractingTwolevel2014,burnettDecoherenceBenchmarkingSuperconducting2019,lisenfeldDecoherenceSpectroscopyIndividual2016,coleQuantitativeEvaluationDefectmodels2010,mullerUnderstandingTwolevelsystemsAmorphous2019}. Other dominant sources of noise are those due to imperfect control and calibration of the qubit \cite{vandijkImpactClassicalControl2019}. These can lead to errors, which take the form of unwanted terms in the Hamiltonian of the idle qubits or driven qubits, and lead to unintended operations \cite{blume-kohoutDemonstrationQubitOperations2017}. 

Detailed noise spectroscopy experiments have found coherent interactions with parasitic two-level system (TLS) defects to be a dominant source of noise in superconducting qubits \cite{coleQuantitativeEvaluationDefectmodels2010}. These defects form part of the environment that the qubit interacts with, and they can lead to not only decoherent errors on the qubits, but also coherent errors if some of the defects are interacting resonantly with the qubits. In this case, when modelling the noise in the qubit, we must account for the non-Markovian dynamics \cite{devegaDynamicsNonMarkovianOpen2017} of the qubit system due to interactions with the environment. In contrast, Markovian noise cannot account for memory effects arising due to the environment. Thus, we need methods to model the non-Markovian noise present in the device.

Hardware-agnostic noise characterisation methods are applicable to various quantum computing platforms, enable cross-platform comparisons, and typically do not require low-level access to the hardware.
Noise in a quantum computer can be characterised in a hardware-agnostic way at different levels: characterisation at the level of algorithms can indicate how good a device is at running certain algorithms \cite{lubinskiApplicationOrientedPerformanceBenchmarks2021, millsApplicationMotivatedHolisticBenchmarking2021}; characterisation at the level of individual gates and quantum circuits can help quantify the performance of individual gates and circuits \cite{nielsenGateSetTomography2021, blume-kohoutTaxonomySmallMarkovian2021}; characterisation at the level of idle qubits can be used to estimate metric such as the qubit relaxation rates \cite{krantzQuantumEngineerGuide2019}. Although these characterisation methods can be useful in estimating the quality and performance of a quantum computer, it can be challenging to relate their results to the physical sources of noise in the device. Also, many of the existing methods for characterisation do not take non-Markovian noise into account. Recently, many attempts have been made at extending existing methods and devising new methods for characterisation of non-Markovian noise \cite{niuLearningNonMarkovianQuantum2019,tancaraKernelbasedQuantumRegressor2022,chenNonMarkovianNoiseCharacterization2020,banchiModellingNonmarkovianQuantum2018,zhangPredictingNonMarkovianSuperconducting2021,white2020demonstration, gulacsiSmokinggunSignaturesNonMarkovianity2023,weiCharacterizingNonMarkovianOffResonant2023}. However, the methods typically have a large overhead in terms of the number of circuits needed for the characterisation.

Recently, the post Markovian master equation \cite{shabaniCompletelyPositivePostMarkovian2005} (PMME) has been used to model non-Markovian dynamics for idle superconducting qubits \cite{zhangPredictingNonMarkovianSuperconducting2021}. However, the methods used to model the dynamics of a time-independent Hamiltonian do not directly generalise to the time-dependent Hamiltonian case, which is found whenever gates are applied on the qubits. For example, the PMME cannot directly be used to model the dynamics of qubits when being driven by  quantum gates. Master equations for driven systems subject to specific kinds of non-Markovian noise have been developed \cite{groszkowskiSimpleMasterEquations2022}, but their applicability to arbitrary gate sequences has not been studied. Modelling the non-Markovian qubit dynamics in the presence of driving allows us to evaluate whether the non-Markovian effects change due to the driving, and also to validate that the model used to generate the idle qubit dynamics corresponds to the physical processes leading to the non-Markovian behaviour. This is because a model that captures the actual physics will be applicable even when the qubit is being driven, while a limited model may only work in certain limits, such as the idle qubit limit. 

Thus, in this work we develop a model for the non-Markovian behaviour of qubits under driving. In contrast to idle qubits, driven qubits have gates repeatedly applied on them. We develop gate sequences composed of mirrored-pseudoidentity gates that allow us to evaluate the non-Markovian behaviour of the qubit when driven by pulses of a range of strengths, and run these gate sequences on an IBM Quantum \cite{ibm_quantum} superconducting qubit system to verify the applicability of the model for real devices. In the obtained results, we use damped oscillations of qubit purity and multi-frequency oscillations in qubit observables as signatures of non-Markovian noise on a single qubit \cite{gulacsiSmokinggunSignaturesNonMarkovianity2023}. We compare the results obtained with non-Markovian models with those of a control model with only Markovian noise. To this aim we fit the noise models to the obtained experimental data and evaluate how closely the models capture the experimental behaviour.

While noise models with a large number of parameters can lead to better fits, they typically achieve this by over-fitting. The evaluation of the parameters for different drive amplitudes allows us to determine if a model is over-fitting: such over-parameterized models lead to unphysical discontinuous and seemingly random changes of the parameters as the drive amplitude is increased, while the parameters in a physically meaningful noise model are required to change continuously. We therefore prevent over-fitting by proposing minimal noise models, which capture the physics of the devices well, while still leading to continuous changes as the drive amplitude is increased. This allows us to systematically evaluate how parameters and their uncertainties evolve when increasing the drive amplitude.

\section{Methodology}
\label{sec:methodology}

\subsection{Non-Markovian noise and its signatures}
\label{subsec:nonmarkovianity}

There are various ways of defining and measuring Markovianity and non-Markovianity; the different definitions are based on concepts including divisibility of the quantum processes, information flow between the system and its environment, and environment memory effects \cite{li2018concepts, chruscinski2014degree,rivas2014quantum}. In this work we use the definition based on the divisibility of the dynamical map corresponding to the evolution of the system. The dynamical map $\mathcal{E}(t_f,t_i)$ operates on the density matrix, $\hat{\rho}$, of a system at time $t_i$, and outputs the density matrix at time $t_f$, which can be expressed as $\hat{\rho}(t_f) = \mathcal{E}(t_f,t_i) [\hat{\rho}(t_i)]$. Using the definition of Markovianity based on divisibility, a process is Markovian if and only if its dynamical map $\mathcal{E}(t_2,t_0)$ can be written as $\mathcal{E}(t_2,t_0)[\cdot] = \mathcal{E}(t_2,t_1)[\mathcal{E}(t_1,t_0)[\cdot]]$ for all times $t_2\geq t_1 \geq t_0$, where $\mathcal{E}(t_j,t_i)$ is a completely positive trace preserving (CPTP) map. If this condition cannot be satisfied, the dynamics is non-Markovian. This is the case, for example, when there is time-correlated dephasing noise on the qubit, or when the qubit is interacting with a strongly coupled TLS.

One signature of non-Markovian noise acting on a qubit can be obtained by analyzing the purity, $p$, of the qubit state, which is defined as $p = \Tr[\hat{\rho}^2]$. The purity is a measure of how mixed a quantum state is: for a single qubit, a purity of $1$ corresponds to a pure state, while a purity of $\frac{1}{2}$ corresponds to the maximally mixed state. If we write a general density matrix of a single qubit as
\begin{equation}
\begin{split}
&\hat{\rho} = \frac{1}{2}\qty(\hat{I} + \Tr[\hat{\rho}\sx]\sx + \Tr[\hat{\rho}\sy]\sy + \Tr[\hat{\rho}\sz]\sz),\\
\end{split}
\end{equation}
its purity is
\begin{equation}
p = \frac{1}{2} \qty(1 + \Tr[\hat{\rho}\sx]^2 + \Tr[\hat{\rho}\sy]^2 + \Tr[\hat{\rho}\sz]^2).
\end{equation}

Generally, the purity of a qubit decreases with time due to decoherent errors such as qubit dephasing. However, non-unital quantum channels such as amplitude damping, which do not preserve the maximally mixed state, can lead to purity increasing with time as well. Oscillations of qubit coherence and purity over time indicate the presence of non-Markovian noise \cite{gulacsiSmokinggunSignaturesNonMarkovianity2023} in idle qubits, because they are a signature of a backflow of information from the environment to the system \cite{rivas2014quantum}. For a driven qubit on the other hand, while oscillations in the purity can be caused by non-Markovianity, they can also be caused by a periodic, time-dependent Hamiltonian and a non-unital decoherent channel. For example, one can generate purity oscillations in the presence of strong amplitude damping by periodically applying an $X$ gate on the qubit. Even in this case, however, the oscillations cannot take an arbitrary form if the noise is purely Markovian. More specifically, in \appref{app:purity_oscillation_proof} we show that oscillations of the qubit purity of the form
\begin{equation}
p(t) = \frac{1}{2} + d(t)\cos^2\qty(f_p t),
\label{eq:pt_decay}
\end{equation}
where $d(t)$ is a non-periodic function of time, $t$, cannot be observed for a qubit undergoing Markovian evolution with repeated applications of a dynamical map $\mathcal{E}$.
For example, as shown in \appref{app:purity_oscillation_proof}, a damped purity oscillation of the above form with $d(t) = e^{-\gamma t}$ ,where $\gamma \neq 0$, cannot be generated upon repeated application of the same dynamical map if the underlying noise process is Markovian. Thus, the presence of such damped oscillations of qubit purity with the number of repetitions of a quantum process is a signature of non-Markovianity for a driven qubit. 

Another signature of non-Markovian dynamics of both idle and driven qubits is the presence of multi-frequency oscillations of qubit observables with the number of repeated gate sequences. In the presence of time-independent Markovian noise on a qubit, a single-frequency, damped oscillation of qubit observables,  such as $\expval{\sx}$, with the number of repetitions of operations is found. The damping arises from decoherent noise effects, while the single-frequency oscillations arise from the unitary operations and coherent noise. In \appref{app:repeated_markovian_process_proof} we show that repeated Markovian processes applied to a single qubit can only lead to damped oscillations with at most a single frequency. Thus, the presence of multi-frequency oscillations in observables such as $\expval{\sx}$ can also act as a signature of non-Markovianity. This has also been proposed in Ref. \cite{gulacsiSmokinggunSignaturesNonMarkovianity2023}. 

\subsection{Noise characterisation method}
\label{subsec:characterisation_method}

Characterisation of coherent errors in qubits typically involves repeatedly applying a sequence of gates in order to amplify the effect of the coherent errors, thereby improving the accuracy and precision of the characterisation, and limiting the contributions of un-amplified errors such as qubit readout errors \cite{vitanovRelationsSingleRepeated2020,nielsenGateSetTomography2021,PhysRevA.87.062119,PRXQuantum.1.020318,magesan2020effective, sheldonProcedureSystematicallyTuning2016}. One prepares the qubit(s) in some initial state, repeatedly applies a chosen gate sequence, and then performs measurement in some basis. Different choices of gate sequences, initial states, and measurement basis can amplify different errors on the qubit, or can make the qubit insensitive to certain errors. By appropriately choosing these parameters, one can create a set of circuits that allows for the characterisation of various error contributions in the qubit operations \cite{nielsenGateSetTomography2021, reed2013entanglement}.

Once the set of noise characterisation circuits have been run on hardware, one needs to analyse the results in order to quantify the significance of the different noise sources. Our method of quantifying the significance of various noise sources involves first developing a minimal noise model which includes the relevant sources of noise present in the device. A minimal model also reduces the chances of over-fitting to the experimental data. We then implement the noise model in our quantum circuit emulator. Finally, we use non-linear regression to fit noise model parameters to experimental data by finding the optimal parameters which lead to the smallest mean-squared-error between the experimental data and the data generated by the noisy emulator. The obtained noise model parameters quantify the significance of the corresponding noise sources.

 In this section, we first describe the mirrored-pseudoidentity error amplifying gate sequence. Next, we show the full set of noise characterisation circuits that we run. Lastly, we describe the non-linear regression method as well as the methods used to analyse the data.
 
\subsubsection{Mirrored-pseudoidentity gate sequence}
\label{subsubsec:mirrored_pseudoidentity_gates}
We define a pseudoidentity operation as one which equates to the identity operation in the absence of noise. However, in the presence of noise, the pseudoidentity operations will in general lead to unitary and non-unitary dynamics on the qubit. 
We want to develop gate sequences that allow amplifying the effects of non-Markovian noise. To achieve this goal, we have developed a mirrored-pseudoidentity operation that is parameterised by the amplitude of the driving pulse. Different driving amplitudes effectively correspond to different gate sequences, and tuning the amplitude of the driving pulse allows observing how the non-Markovian effects are affected by the different gate sequences.

Let $X_\theta$ represent a potentially noisy gate acting on a qubit that performs a rotation about the $\xax$ axis in the Bloch sphere by an angle $\theta$. In the absence of any noise, the unitary operator describing this gate is given by 
\begin{equation}
\hat{X}(\theta) = e^{-i\theta \sx/2},
\end{equation}
where $\sx$ is the Pauli-$X$ operator. Note that we are using $X_\theta$ to denote the gate and the operator $\hat{X}(\theta)$ to denote the unitary corresponding to the gate in the absence of noise. $Y_\theta, Z_\theta, \hat{Y}(\theta)$, and $\hat{Z}(\theta)$ are defined analogously. Our mirrored-pseudoidentity gate, $U_\theta$, is composed of an $X_\theta$ gate followed by an $X_{-\theta}$ gate. In the absence of noise, the unitary operator is equal to the identity operation:
\begin{equation}
\hat{U}(\theta) = \hat{X}(-\theta) \hat{X}(\theta) = \hat{I}.
\end{equation}
Note that for the case $\theta = 0$, we are applying a $0$ amplitude pulse, which corresponds to an ``idle" operation in the absence of any noise with the same duration as that of a single qubit gate. 

We choose this gate sequence because over-rotation errors of the form $\hat{X}(\theta) \rightarrow \hat{X}(\theta(1+\epsilon))$ are cancelled to first order in $\epsilon$ for all $\theta$, while noise of the form
\begin{equation}
e^{-i\frac{\theta}{2} \sx} \rightarrow e^{-i(\frac{\theta}{2} \sx + \delta \omega \sz)},
\end{equation}
where $\delta \omega$ denotes the ``phase error", is not cancelled in general. Thus, this gate sequence will amplify phase errors but not over-rotation errors in the qubit operations.

\subsubsection{Noise characterisation circuits}
\label{subsubsec:noise_characterisation_circuits}
Having described the pseudoidentity operations that we use, we now describe and explain our choice of initial state. 
Spin-locking measurements \cite{yanRotatingframeRelaxationNoise2013,abdurakhimovDrivenstateRelaxationCoupled2020} involve applying operations on a qubit that rotate it about an axis parallel to the qubit state in the Bloch sphere. This ``locks" the qubit in a specific state even in the presence of external perturbations. On the other extreme, 
significant non-Markovian effects have been shown to occur in the dynamics of idle qubits prepared in the equator of the Bloch sphere \cite{zhangPredictingNonMarkovianSuperconducting2021}. 

In our noise amplifying circuits we can continuously go from one to the other of these extremes by preparing the qubit in the initial state $\ket{+}$: for pseudoidentity $U_0$, we obtain idle qubit dynamics which can show significant non-Markovian noise, and for $U_{2\pi}$ we obtain the robust spin-locking dynamics. 
In \appref{app:hamiltonian_error_insensitivity}, we show that for the pseudoidentity $U_{2\pi}$ acting on an initial state $\ket{+}$, the qubit dynamics becomes insensitive to small, time-independent errors in the Hamiltonian of the qubit of form $\h_\mathrm{ideal}(t) \rightarrow \h_\mathrm{ideal}(t) + \epsilon\h_\mathrm{noise}$. 

After the initial state preparation in the $\ket{+}$ state, the pseudoidentity gate sequence is applied $n$ times, followed by measurement of the qubit in the $X, Y,$ and $Z$ basis. This is because the three measurement results correspond to $\expval{\sx},\expval{\sy}$, and $\expval{\sz}$, and they are sufficient to reconstruct the density matrix of the qubit, since $\hat{\rho} = (\mathds{I} + \expval{\sx}\sx+ \expval{\sy}\sy+ \expval{\sz}\sz)/2$. In the absence of any noise, this state tomography would allow reconstructing the initially prepared state, which is $\hat{\rho} = \ket{+}\bra{+}$.

We now discuss the physical implementation of these circuits. In the qubit rotating frame, the Hamiltonian corresponding to a pulse that leads to a rotation about an axis in the $\xax-\yax$ plane of the Bloch sphere  is given by
\begin{equation}
\h (t) = \Omega(t) [\cos(\gamma)\sx + \sin(\gamma)\sy],
\label{eq:general_xy_pulse_hamiltonian}
\end{equation}
where $\Omega(t)$ describes the shape of the pulse envelope, and $\gamma$ controls the axis of rotation. The amount of rotation in the Bloch sphere depends on the integral of the pulse envelope $\Omega(t)$ over the duration of the gate. Thus, the amount of applied rotation depends approximately linearly on the pulse amplitude. This is an approximation due to errors not considered here such as the finite rise time of the pulses \cite{SangchulPhysRevB.65.144526} or non-linearities in the qubit drive line components \cite{lazarCalibrationDriveNonLinearity2022}.
Thus, we implement an $X_\theta$ operation for different values of $\theta$ in a single pulse by changing the amplitude of the applied pulse. Let $\Omega_{X_{\pi / 2}}(t)$ represent the pre-calibrated pulse envelope corresponding to the $X_{\pi / 2}$ gate.  In the qubit rotating frame, the Hamiltonian for this operation is 
\begin{equation}
\h_{X_{\pi / 2}} (t) = \Omega_{X_{\pi / 2}}(t) \sx.
\end{equation}
We then implement a single-pulse $X_{\theta}$  gate by linearly scaling the amplitude $\Omega$ of the pulse, which gives the following Hamiltonian:
\begin{equation}
\h_{X_{\theta}}(t) = \frac{\theta}{\pi/2} \h_{X_{\pi / 2}} (t).
\end{equation}
We implement negative rotations $X_{-\theta}$ by applying the same $X_{\theta}$ operation, but with an additional virtual $Z_\pi$ gate \cite{mckayEfficientZGatesQuantum2017} before and after each $X_{\theta}$. Virtual Z gates apply a phase change to all future pulses. The 
$Z_\pi$ gate leads to $\gamma \rightarrow \gamma + \pi$ for the pulses between the $Z$ gates. The transformation $\gamma \rightarrow \gamma + \pi$ leads to $ \h_{X_{\theta}}(t) \rightarrow  - \h_{X_{\theta}}(t) $, which in turn leads to $ X_\theta \rightarrow X_{-\theta}$. 

Non-linearity of qubit drive line components can lead to the amplitude of the pulse reaching the qubit not varying strictly linearly with the amplitude of the generated pulse \cite{lazarCalibrationDriveNonLinearity2022}. This can lead to over/under-rotations when applying $X_\theta$ gates with $\theta \neq \pi/2$. As discussed earlier, our choice of gate sequences does not amplify over-rotation errors of this form. Hence, our results are not significantly affected by the qubit drive line non-linearity.
Although such over-rotation errors are not amplified in our gate sequence, we still avoid applying pulses with very large amplitudes, since these lead to additional sources of noise, and potentially break the symmetry between $\h_{X(\theta)}(t)$ and $-\h_{X(\theta)}(t)$, which is required for the cancellation of the over-rotation errors. 

\begin{figure}[t!]
  \vspace*{0.15in}
  \hspace*{0in}
  \Qcircuit @C=1em @R=.9em {
    & & \mbox{\hspace{19em}$\cross n$} & & \\
    & \mbox{\quad\quad\quad\quad$\cross m$} & & \mbox{\quad\quad\quad\quad$\cross m$} &  \\
    \lstick{\ket{+}} &  \gate{X_{\thetagate}}  & \gate{Z_\pi}& \gate{X_{\thetagate}}  & \gate{Z_\pi} & \measure{\mbox{X/Y/Z}}    \gategroup{3}{2}{3}{2}{.5em}{.}\gategroup{3}{4}{3}{4}{.5em}{.}\gategroup{2}{2}{3}{5}{1.0em}{--}
    }
\caption{Quantum circuit for the noise characterisation; after preparation of the qubit in the $\ket{+}$ state, the gates inside the outer dashed box, corresponding to the pseudoidentity gate $U(\thetafull= m\;\thetagate)$, are repeated $n$ times before measurement in the $X,Y$, or $Z$ basis. The initial state preparation and the measurement basis changes are done using additional individual $X_{\pi/2}$ gates and virtual-$Z$ gates.}
\label{fig:circuit_for_repeated_pseudoidentity}
\end{figure}
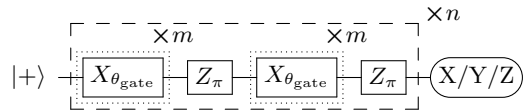

To be able to achieve large rotation angles without using large pulse amplitudes, we implement the $X_\theta$ gates in the full pseudoidentity gate - hereon denoted $X_{\theta_ \text{full}}$ - by evenly dividing them into $m$ smaller, consecutive rotations, which each have the Hamiltonian $\h_{X_{\theta_\mathrm{gate}}}(t)$, where $\thetagate = \thetafull/m$. \figref{fig:circuit_for_repeated_pseudoidentity} shows the full circuit diagram corresponding to our error amplification protocol. The choice of the value of $m$ depends on the maximum value of $\thetafull$ that one aims to run, and on the largest single pulse rotation that can be accurately implemented on the hardware. 

When running experiments on the device, it is possible that two identical batches of circuits, which are sent to the device even shortly after each other, can give very different results due to changes in the device noise properties between the batches. On the other hand, circuits sent in the same batch have a lower probability of exhibiting different results. For example, if the time taken between executions of different batches of circuits is much larger than the time taken between individual quantum circuits within a batch, the noise properties of the device can change significantly between the batches, but not between the circuits within a batch \cite{carrollDynamicsSuperconductingQubit2022, wilson2020jitnoise}. Thus, we send as many circuits as possible in a single batch to minimise risk of the device properties changing during a single set of runs. 

Instead of running circuits corresponding to all values of $n$, i.e. $n = \{0,1,2,..,n_\mathrm{max}\}$, we run circuits corresponding to evenly spaced multiples of a fixed spacing, $\Delta n$, so that  $n = \{0, \Delta n, 2\Delta n, ... , n_\mathrm{max}\}$. The results for intermediate values of $n$ are interpolated using a cubic spline interpolation method \cite{de1978practical}. This allows significantly reducing the number of circuits required, while still providing enough information for the fitting procedure to accurately obtain the model parameters. The optimal value of $\Delta n$ and $n_\mathrm{max}$ depends on how much noise there is in the device: $n_{\mathrm{max}}$ must be large enough to allow for significant oscillations/decays to be observed, while $\Delta n$ must be small enough to avoid aliasing effects and inaccurate interpolation. We choose these values by performing a number of initial experiments to determine the relevant time scales of the qubit dynamics. 

For each value of $n$ we execute $3$ different quantum circuits corresponding to measurements in the $X/Y/Z$ basis in order to perform state tomography. Performing a state tomography gives us the maximum available information to use in the model fitting procedure, and also allows calculating metrics such as qubit purity.

\subsubsection{Non-linear regression and uncertainty quantification}
\label{subsubsec:non_linear_regression_data_analysis}

The observable dynamics generated by the noise models are in general damped oscillations. Thus, we perform non-linear regression to fit noise models to the obtained experimental data. 
The loss function $\mathscr{L}$ that we minimise during the regression is of form
\begin{equation}
\begin{split}
\mathscr{L}(\boldsymbol{x},\thetafull) = \sum_{n=0}^{n_{\mathrm{max}}} \sum_{b = \{x,y,z\}} \bigl[& \expval{\hat{\sigma}_{b, \mathrm{exp}}(n,\thetafull)} \\
 & - \expval{\hat{\sigma}_{b, \mathrm{sim}}(\boldsymbol{x}, n, \thetafull)}\bigr]^2,
\end{split}
\end{equation}
where $\hat{\sigma}_{b, \mathrm{exp}}$  and $\hat{\sigma}_{b, \mathrm{sim}}$  denote the experimental result and the noisy emulator result, respectively, obtained for measurements in the basis $b$, and the vector $\boldsymbol{x}$ corresponds to the noise model parameters. This loss function sums the squares of the Euclidean distance between Bloch vectors corresponding to the experimental and simulated qubit states for different values of $n$. The quality of the fitting is determined by the root-mean-squared-error (RMSE), given by $\mathrm{RMSE}=\sqrt{\mathscr{L}(\boldsymbol{x}_\mathrm{opt},\thetafull)/3(n_{\mathrm{max}}+1)}$.

As mentioned in the previous section, it is possible for the noise environment to change significantly between executions of two batches of circuits. Thus, only noise model parameters obtained from the results of a single batch of circuits can be expected to be approximately constant across the circuit executions, while parameters fitted within different batches will in general be different due the potentially occurring changes in the device properties between one batch and the next. To investigate the changes of parameters as function of $\thetafull$, all the circuits used to fit those parameters at different values of $\thetafull$ need to be executed within one batch. 

When evaluating the changes of the noise parameters with increasing $\thetafull$, one can perform the fit for each angle $\thetafull$ independently. However, some parameters will be largely independent of the value of $\thetafull$, while others may exhibit larger changes. Having additional parameters in the model can lead to over-fitting, where the model attempts to account for the inherent disturbances or fluctuations present in the experimental data itself, such as the variability introduced by the finite number of shots used to estimate the qubit observables. To minimize the amount of over-fitting, it can be advantageous to fix the parameters that are approximately independent of $\thetafull$ to a single value identical for all $\thetafull$ considered within one batch of runs. To determine which parameters are approximately constant, one can perform an initial fit, where the parameters of each $\thetafull$ are fitted independently, and from this determine which are the ones that do not change significantly. These can then be fixed in a subsequent fit, while the parameters that change significantly with $\thetafull$ are allowed to change.
To evaluate how the noise model parameters change over different $\thetafull$, where some parameters are kept constant, we minimise the loss function corresponding to the sum of loss functions for the set of considered $\thetafull$, 
so that $\mathscr{L}_{\mathrm{total}} = \sum_{\thetafull} \mathscr{L}(\boldsymbol{x}(\thetafull),\thetafull)$, 
where we have explicitly denoted that the parameters $\boldsymbol{x}$ are functions of $\thetafull$. 
When keeping some noise model parameters constant for different $\thetafull$, one needs to ensure that the total loss function is not significantly changed compared to the case where one allows all parameters to change for different $\thetafull$.

While the noise parameters themselves can change significantly between different batches of circuits, the relative changes of parameters can be expected to vary less. In each batch that we run, we therefore execute circuits for both a finite $\thetafull$ and for $\thetafull = 0$. We then calculate the ratio of the noise parameters $r_{i}(\thetafull) = \frac{x_i(\thetafull)}{x_i(0)}$. Rather than comparing the changes of the parameters themselves as function of $\thetafull$ we therefore compare changes of this ratio $r_{i}(\thetafull)$.

For some angles $\thetafull$ there can be a significant uncertainty in the fit of specific individual parameters. This is the case when a large change in a specific parameter for a given $\thetafull$ only gives very small changes in the loss function. For such parameters, the fitted values as function of theta will fluctuate significantly due to the large uncertainty. It is therefore important to compute the uncertainty of the fit, since it allows distinguishing physically driven large fluctuations of parameters from spurious ones that originate from large fitting uncertainty.
We therefore need to estimate the uncertainty in $r_{i}(\thetafull)$, which we denote as $\sigma_{i}(\thetafull)$, that results from the non-linear regression. We describe the procedure in \appref{app:uncertainty_estimation}. 

We run the experiments corresponding to all angles in the set $\{\thetafull\}$ every day for $K$ days, which gives $K$ estimates of each $r_{i}(\thetafull)$ and $\sigma_{i}(\thetafull)$. The next step is to aggregate the $K$ results obtained for the same experiments run at different times and account for the additional uncertainty due to the distribution of results. Let $r_{ik}(\thetafull)$ denote the $k$-th estimate of the ratio $r_{i}(\thetafull)$.  In order to aggregate the results for different values of $k$ while de-weighing the effects of results with large uncertainties, we compute the weighted mean \cite{cochran1977sampling}, $\tilde{r}_{i}(\thetafull)$, of $r_{ik}(\thetafull)$ with the weights being given by the inverse of the corresponding variances. Thus,
\begin{equation}
\tilde{r}_{i}(\thetafull) = \frac{1 }{\sum_k \frac{1}{\sigma_{ik}(\thetafull)^{2}}}\sum_k \frac{r_{ik}(\thetafull) }{\sigma_{ik}(\thetafull)^{2}}.
\end{equation}

In order to estimate the total uncertainty associated with $\tilde{r}_{i}(\thetafull)$, we first need to evaluate the fitting uncertainty due to $\sigma_{ik}(\thetafull)$. This is given by
\begin{equation}
\tilde{\sigma}^\mathrm{f}_{i}(\thetafull) = \frac{1}{\sqrt{\sum_k \frac{1}{\sigma_{ik}(\thetafull)^{2}}}}.
\end{equation}
The second source of uncertainty is due to the distribution of $r_{ik}(\thetafull)$ across different days. Instead of using standard deviation, which would give the same weight to all data points, even those with high uncertainty, we calculate the weighted standard deviation \cite{cochran1977sampling}, which is given by
\begin{equation}
\tilde{\sigma}^\mathrm{d}_{i}(\thetafull) = \sqrt{\frac{1 }{\sum_k \frac{1}{\sigma_{ik}(\thetafull)^{2}}}\sum_k \frac{[r_{ik}(\thetafull) -
\tilde{r}_{i}(\thetafull)]^2 }{\sigma_{ik}(\thetafull)^{2}}}.
\end{equation}
Finally, the total uncertainty \cite{farrance2012uncertainty} in $\tilde{r}_{i}(\thetafull)$ is given by
\begin{equation}
\tilde{\sigma}_{i}^{\mathrm{total}}(\thetafull) = \sqrt{\tilde{\sigma}_{i}^\mathrm{f}(\thetafull)^2 + \tilde{\sigma}_{i}^\mathrm{d}(\thetafull)^2}.
\end{equation}

\subsection{Noise models}
\label{subsec:noise_models}

The models of the effects of noise in superconducting qubits are typically based on different levels of assumptions \cite{jonesApproximationsTransmonSimulation2021}. Making fewer assumptions allows for higher accuracy simulations, but at the cost of increased simulation time and complexity. In this work, the approximations we make include limiting simulations to the qubit subspace and thereby ignoring presence of higher energy levels, neglecting the effects of the pulse envelope shapes by approximating the Hamiltonian for corresponding to a gate as constant, and omitting the effects of state-preparation and measurement (SPAM) errors. We neglect SPAM errors in our model because they don't get amplified by the repeated pseudoidentities. These approximations, combined with the finite number of noise sources included in our noise models, will lead to a remaining discrepancy between the experimental results and the predictions of the noise model after the non-linear regression. The amount of discrepancy gives an indication of the validity of the approximations for the considered hardware.

Below, we present three different noise models, where each of them deals with non-Markovianity in a different way: \begin{enumerate*}
    \item model does not include non-Markovian noise; 
    \item model incorporates non-Markovian noise by allowing the qubit to interact with another effective two-level quantum system;
    \item model incorporates non-Markovian noise by a phenomenological memory kernel used in the post Markovian master equation.
\end{enumerate*}
Including a Markovian model allow us to assess how significant the non-Markovian noise effects are. We choose two different approaches for modelling non-Markovianity in order to compare how well different approaches work in modelling the noise seen on hardware, and also to analyse the relation between the different approaches. Furthermore, if consistency in results is found across equivalent noise models, it serves as a validation for the non-linear regression and the model parameters obtained, and adds trustworthiness in the fitted data.

\subsubsection{Markovian noise model}

Our first model is a model that doesn't contain any non-Markovian components; we denote this as the ``Markovian model". If the qubit hardware exhibits non-Markovian noise, this model will be unable to fit the experimental data in general. It also provides a foundation for the other noise models, which add non-Markovian components on top of this model. 

The dominant decoherent noises in superconducting qubits available today can be modelled as amplitude damping and pure qubit dephasing \cite{krantzQuantumEngineerGuide2019}. Dominant coherent errors include phase errors, caused by shift in qubit frequencies or by application of pulses, and over-rotation errors, which can be caused by miscalibration of the gates. By design, our choice of pseudoidentity gate is insensitive to over-rotation errors, which allows us to reduce the number of parameters in the noise model by omitting over-rotation errors.  

The qubit dynamics is simulated using the Lindblad Master equation
\begin{align}
\dot{\hat{\rho}}(t) = \lin [\hat{\rho}(t)] =&  -i[\h(t),\hat{\rho}(t)] + \\ & \sum_{i} \Gamma_i \qty(\lhat_i\hat{\rho}(t)\lhat_i^\dagger - \frac{1}{2}\qty{\lhat_i^\dagger \lhat_i, \hat{\rho}(t)}),
\label{eq:lindblad_master_equation}
\end{align}
where $\lin$ is the Lindbladian generator of the dynamics, $\h(t)$ includes the unitary evolution components, and the Lindblad operators $\lhat_i$ and their corresponding coefficients $\Gamma_i$ lead to decoherent evolution.

The phase error modifies the ideal Hamiltonian as follows: 
\begin{equation}
\h(t) = \h_\text{ideal}(t) + \delta \omega \hspace{0.05cm}  \sz,
\end{equation}
where $\delta \omega$ quantifies the amount of phase error. The time-dependence in the Hamiltonian comes from the application of different gates in the quantum circuit. Since we are not taking pulse shapes into account, each gate is modelled as a time-independent Hamiltonian acting on the qubit for a fixed duration.

In this model, the summation in \eqref{eq:lindblad_master_equation} goes over the the Lindblad operators corresponding to amplitude damping and dephasing. The amplitude damping noise channel is defined by the Lindblad operator \cite{lindblad1976generators}, which in matrix form is given by
\begin{equation}
L_{AD} = \left(
\begin{array}{cc}
 0 & 1\\
  0&0
  \end{array}
  \right),
\label{eq:amplitude_damping_lindblad_operator}
\end{equation}
and the dephasing noise channel, which is given by
\begin{equation}
L_{D} = \left(
\begin{array}{cc}
 1 & 0\\
  0&-1
  \end{array}
  \right).
\end{equation}
Thus, in this model, the real numbers $\Gamma_{AD}$ and $\Gamma_{D}$ quantify the strengths of the amplitude damping and dephasing channels, respectively.

\subsubsection{Qubit-TLS model}
\label{sssec:Qubit-TLSmodel}

The second noise model introduces non-Markovianity by allowing the qubit to interact with another, resonant quantum system. We choose the additional quantum system to be a TLS. This is a minimal model for strong interactions with the environment. Such interactions have been observed experimentally in various forms, for instance, in nearly-on-resonant qubit-defect interactions in superconducting qubits  \cite{klimovFluctuationsEnergyRelaxationTimes2018, schlorCorrelatingDecoherenceTransmon2019,zhaoCombatingFluctuationsRelaxation2022,carrollDynamicsSuperconductingQubit2022,burnettDecoherenceBenchmarkingSuperconducting2019}. Note that dynamics of the combined Hilbert space of the qubit and the TLS is Markovian, and we only get non-Markovian dynamics on the qubit subsystem after tracing out the TLS. 

Motivated by existing experimental results in literature \cite{zhangPredictingNonMarkovianSuperconducting2021}, we consider the case where non-Markovian dynamics of an idle qubit is visible when the qubit is prepared in a state on the equator of the Bloch sphere, but not when the qubit is prepared in the $\ket{0}$ or $\ket{1}$ state. More specifically, the non-Markovian interaction between the qubit and the TLS leaves the $\ket{0}$ and $\ket{1}$ states of the qubit unaffected. We can use a $\sz$ operation on the qubit to model the effect of the qubit-TLS interaction on the qubit. 

Let $\hat{\tau}_i$ denote Pauli operators acting on the TLS. The minimal model for the qubit-TLS interaction that can lead to non-Markovian effects consists of the full Hamiltonian $\h_\mathrm{full} = \h_\mathrm{qubit}\otimes\hat{I} +  \nu_{zj}\sz\otimes\hat{\tau}_j$, where $\hat{\tau}_j$ is an arbitrary Pauli operator. 
Let $\ket{\hat{\tau}_j^\pm}$ represent the eigenstates of $\hat{\tau}_j$ with eigenvalues $\pm 1$, respectively. We can write the initial state of the TLS in this basis as
\begin{equation}
\ket{\phi_\mathrm{TLS}(0)} = c^+_\mathrm{TLS}(0)\ket{\hat{\tau}_j^+} + c^-_\mathrm{TLS}(0)\ket{\hat{\tau}_j^-},
\end{equation}
where $c^\pm_\mathrm{TLS}(0)$ are the coefficients corresponding to the eigenstates $\ket{\hat{\tau}_j^\pm}$. The chosen form of the qubit-TLS interaction leads to a qubit frequency dependence on the TLS state. The effective qubit Hamiltonian when the TLS is in the state $\ket{\hat{\tau}_j^\pm}$ is given by
\begin{equation}
\h_\mathrm{qubit}^\pm = \h_\mathrm{qubit} \pm \nu_{zj}\sz.
\end{equation}

The $\nu_{zj}\sz\otimes\hat{\tau}_j$ term in the Hamiltonian only affects the complex phase of $c^\pm_\mathrm{TLS}(0)$ and not their magnitude. Thus, the probability of measuring the TLS in the $\ket{\hat{\tau}_j^\pm}$ states is unchanged by the interaction. Therefore, the qubit dynamics is only affected by the value of $|c^+_\mathrm{TLS}(0)|^2 = 1 - |c^-_\mathrm{TLS}(0)|^2$. The choice of $|c^+_\mathrm{TLS}(0)|^2 = \frac{1}{2}$ corresponds to an equal probability of the two effective qubit frequencies. Note that $|c^+_\mathrm{TLS}(0)|^2 = 0$ or $|c^+_\mathrm{TLS}(0)|^2 = 1$ would lead to Markovian qubit dynamics and the error would be indistinguishable from the phase errors of the form $\delta \omega \sz$.

In our minimal model for non-Markovian dynamics, we set $|c^+_\mathrm{TLS}(0)|^2 = \frac{1}{2}$ by setting $\hat{\tau}_j = \hat{\tau}_x$ and setting the initial TLS state to be $\ket{0}$. In a more general model one can also allow for an asymmetry between the two effective qubit frequencies by having a qubit TLS interaction of form $\nu_{zx}\sz\otimes\hat{\tau}_x + \nu_{zz}\sz\otimes\hat{\tau}_z$ and letting the TLS state be $\ket{0}$, or by letting the interaction remain as $\nu_{zx}\sz\otimes\hat{\tau}_x$, while having an initial general TLS state of the form $c^+\ket{0} + c^-\ket{1}$.
Furthermore, one can also include terms in the idle TLS Hamiltonian of form $\epsilon_x \hat{\tau}_x + \epsilon_z \hat{\tau}_z $, consider time-dependent Hamiltonians that arise due to change in rotating frames, and consider the presence of multiple interacting TLSs \cite{PhysRevB.91.014201}. However, when using such extended models in our fits to experimental data for the considered device, we find that they lead to over-parametrisation and hence over-fitting, while only marginally reducing the RMSE of the fit when compared to our minimal qubit-TLS model.

We note here that the qubit-TLS model can also describe the non-Markovian dynamics caused by other physical mechanisms, such as qubit frequency fluctuations caused by trapped quasiparticles or incoherent TLSs \cite{riste2013millisecond, deGraafTwoLevelSystemsInSuperconducting, PhysRevB.103.174103, thorbeck2022tls}. When such noise sources are present the qubit frequency can fluctuate between two values. If the fluctuation time is longer than an individual quantum circuit duration, but shorter than the averaging time over the multiple shots, then the model presented above can describe the resulting qubit dynamics, since the qubit-TLS interaction effectively leads to an equivalent  mixture of qubit frequencies shifted by the qubit-TLS interaction strength.

We introduce dissipation on the TLS by including an amplitude damping channel with strength $\kappa$ on the TLS. The Lindblad operator on the TLS subspace is identical to the qubit amplitude damping operator shown in \eqref{eq:amplitude_damping_lindblad_operator}. The qubit-TLS model is identical to the Markovian model, except for the addition of the interaction $\nu_{zx}\sz\hat{\tau}_x$ with a TLS, and amplitude damping on that TLS. Based on the discussion in the preceding paragraphs, to simulate this interaction we need to effectively model the TLS as a second qubit, where the governing Hamiltonian of the full system is given by
\begin{equation}
\h(t) = (\h_\text{ideal}(t) + \delta \omega \sz)\otimes \hat{I} + \nu_{zx}\sz \otimes \hat{\tau}_x,
\label{eq:qubit_tls_model_hamiltonian_equation}
\end{equation}
and the governing Lindblad operators change as $\lhat_i \rightarrow \lhat_i\otimes \hat{I}$ for the qubit and as $\lhat_i \rightarrow \hat{I}\otimes \lhat_i $ for the TLS.

In \appref{app:undriven_qubit_tls_solution} we derive an equation that describes the dynamics of such a system for the undriven case in the absence of amplitude damping and qubit dephasing, and with the  qubit and TLS initially prepared in the $\ket{+}$ and $\ket{0}$ state, respectively. The density matrix of the qubit subsystem in this case is given by
 \begin{equation}
\begin{split}
    \hat{\rho}(t)&= \frac{1}{2}\left(\begin{array}{cc} 1 & e^{-2i\delta\omega t} \cos(2\nu_{zx}t) \\ e^{2i\delta\omega t} \cos(2\nu_{zx}t) & 1 \end{array}\right).
\end{split}
\end{equation}
The purity of the qubit is then given by
\begin{equation}
p(t) = \Tr[\hat{\rho}(t)^2] = \frac{1}{2} + \frac{1}{2}\cos(2\nu_{zx}t)^2.
\end{equation}
This confirms that qubit-TLS model describes oscillations of the qubit purity for the idle qubit, which are a signature of non-Markovianity, as discussed in \secref{subsec:nonmarkovianity}. 

\subsubsection{Post Markovian master equation model}
Finally, our third noise model uses the post Markovian master equation (PMME) \cite{shabaniCompletelyPositivePostMarkovian2005} to model non-Markovian noise effects. The PMME has previously been used to model non-Markovian noise in idle superconducting qubits \cite{zhangPredictingNonMarkovianSuperconducting2021}, and we use it here to compare its results for the idle qubit with those of the other two noise models. The PMME is written as
\begin{equation}
  \begin{split}
\dv{}{t} \hat{\rho}(t) &= \lin _0 \hat{\rho}(t) + \lin_1 \int_0^t dt'k(t') e^{(\lin_0 + \lin_1)t'} \hat{\rho}(t-t'),\\
\label{eq:orig_pmme}
\end{split}
\end{equation}
where $\lin_0, \lin_1$ are the Lindbladian generators corresponding to the Markovian and to the non-Markovian evolution, respectively. These operators correspond to physical processes, and have the form shown in \eqref{eq:lindblad_master_equation}. Non-Markovian effects under the evolution of $\lin_1$ are introduced by the use of a phenomenological memory kernel $k(t)$, which assigns weights to historical system states. This memory kernel is phenomenological because it is used to interpolate between Markovian and exact dynamics in the measurement interpretation \cite{shabaniCompletelyPositivePostMarkovian2005}. It is only equal to the memory functions which relate to the spectral density of the reservoir, such as those used in the second order Nakajima-Zwanzig equation \cite{nakajima1958quantum,zwanzig1960ensemble}, in certain limits such as the weak coupling limit \cite{devegaDynamicsNonMarkovianOpen2017, maniscalcoNonMarkovianDynamicsQubit2006}. Note that in the limit $k(t) \rightarrow \delta(t)$, we obtain Markovian dynamics. 
In this article we use this model for the undriven qubit case, the extension of the PMME to time-dependent Hamiltonians will be presented separately as part of future work.

In addition to the sources of noise in the Markovian model, in the PMME model, following Ref. \cite{zhangPredictingNonMarkovianSuperconducting2021}, we include a non-Markovian dephasing noise channel. The Lindbladian generators in the PMME model are described below:
\begin{itemize}
    \item $\lin_0$ is identical to the Markovian model Lindbladian, and includes the effect of a phase error, amplitude damping and dephasing.
    \item $\lin_1$ is the Lindbladian corresponding to the non-Markovian dephasing, and is given by $\lin_1 [\hat{\rho}(t)]=  \gamma_z \qty(\sz\hat{\rho}(t)\sz - \hat{\rho}(t))$, where $\gamma_z$ quantifies the amount non-Markovian dephasing. 
\end{itemize}

We use time units in which the duration of a single gate is $1$, leading to the memory kernel and the noise model parameters being dimensionless. The memory kernel $k(t)$ that we use is parameterised by the memory kernel decay rate $b$, and has the form $k(t) = e^{-bt}$. Such a memory kernel has been used in Ref. \cite{shabaniCompletelyPositivePostMarkovian2005, zhangPredictingNonMarkovianSuperconducting2021}. Substituting our chosen memory kernel into \eqref{eq:orig_pmme} leads to the PMME master equation
\begin{equation}
\label{eq:zhang_paper_pmme}
  \begin{split}
\dv{}{t} \hat{\rho}(t) = \lin _0 \hat{\rho}(t) + \lin_1 \int_0^t dt'e^{-bt'} e^{(\lin_0 + \lin_1)t'} \hat{\rho}(t-t').
\end{split}
\end{equation}
We note that more general forms of the memory kernel can be used, which allow modelling a wider range of non-Markovian dynamics.

\subsubsection{Relations between the noise models}
\label{subsec:PMME_qtls_relation}
Below we summarise the free parameters in each of the three noise models described above.
\begin{enumerate}    
  \item \makebox[3cm]{Markovian model: \hfill}$\delta \omega, \Gamma_{AD}, \Gamma_{D}$.
  \item \makebox[3cm]{Qubit-TLS model: \hfill}$\delta \omega, \Gamma_{AD}, \Gamma_{D}, \nu_{zx}, \kappa$.
  \item \makebox[3cm]{PMME model: \hfill}$\delta \omega, \Gamma_{AD}, \Gamma_{D}, \gamma_z, b$.
\end{enumerate}
We note that all the noise models have three parameters that determine the Markovian noise contributions. Both the qubit-TLS and the PMME noise models add two further parameters to describe the non-Markovian components of the dynamics. The mathematical operators corresponding to the noise sources denoted by $\delta \omega$, $\Gamma_{AD}$, and $\Gamma_{D},$ are the same for the three noise models, and they describe the same physics. Thus, the values obtained for the corresponding noise terms by non-linear regression should approximately be the same for all three noise models. The method for the noisy simulation of the Markovian and Qubit-TLS model is described in \appref{app:markovian_me_simulation}. Simulations for the PMME model are done by semi-analytically solving the Master equation \eqref{eq:zhang_paper_pmme}.

We now demonstrate that the parameters in the PMME model are directly related to those in the qubit-TLS model. In \appref{app:pmme_zx_interaction} we present the general analytical solution of the PMME and of the qubit-TLS model for the idle qubit. In that derivation we neglect the amplitude damping channel acting on the qubit in both models for increased clarity of the resulting equations. At the end of \appref{app:pmme_zx_interaction} in \eqref{eq:qubit_tls_pmme_parameter_relations} we show that the parameters of the two models can be related exactly with the following relations:
\begin{align}
\gamma_z= &  2\nu_{zx}^2,\label{eq:gammaz_nuzx}\\
b =& \frac{\kappa}{2} - 4\nu_{zx}^2.
\label{eq:main_text_pmme_qtls_relation}
\end{align}
These equations show that $\gamma_z$ corresponds to twice the interaction strength $\nu_{zx}$ between the qubit and the TLS squared. The memory kernel decay rate $b$ relates to the dissipative term on the TLS,  $\kappa$, but also includes a contribution from the qubit-TLS interaction strength with a negative sign. 

In the derivation of the PMME, in the measurement picture, one assumes that after a measurement of the bath the evolution of the system is approximately Markovian \cite{shabaniCompletelyPositivePostMarkovian2005}. This assumption, which corresponds to first-order perturbation theory, is only valid in the limit of the characteristic time of the bath memory effects, given by $\frac{1}{\kappa}$, being must shorter than the characteristic time of scale of the system-environment interaction, given by $\frac{1}{\nu_{zx}}$. In the qubit-TLS model, this assumption corresponds to $\kappa > \nu_{zx}$. In the absence of any decoherent effects on the bath, where $\kappa = 0$, the perturbation theory assumption in the derivation of the PMME becomes invalid \cite{devegaDynamicsNonMarkovianOpen2017}. In fact, with Eq. \ref{eq:main_text_pmme_qtls_relation} one can see that in this case the value of $b$ becomes negative. A negative valued $b$ leads to unbounded, exponentially growing memory kernels towards earlier times. These have previously been found in the case of exceptional points in the generator of reduced system dynamics \cite{ng2022long}. For a detailed discussion on the validity of the PMME, we refer the reader to Ref. \cite{maniscalcoNonMarkovianDynamicsQubit2006}, where the authors compare the PMME to the exact dynamics of a system containing a qubit interacting with a Bosonic reservoir.

\section{Results}

\subsection{Implementation on IBM Quantum hardware}
\label{subsec:implementation_on_IBMQ}
In this section, we discuss the implementation of the noise characterisation procedure on the IBM Quantum device called ``\textit{ibmq\_armonk}" \cite{ibm_quantum}, which is a single qubit device that allows for pulse-level control of the gates \cite{Qiskit}. Note that \textit{ibmq\_armonk} device was retired in July 2022.

The pre-calibrated pulse on the hardware is the pulse for the $X_{\pi/2}$ gate. Gates for general $X_\theta$ don't need to be calibrated, because any arbitrary single qubit gate can be constructed using $X_{\pi/2}$ and $Z_{\theta}$ gates. The pulse shape used is a so called DRAG pulse \cite{motzoiSimplePulsesElimination2009}, which is designed to reduce leakage outside the qubit subspace. Each single qubit gate has a duration of $71.1\  \unit{\nano\second}$. We choose $m = 4$, thus dividing each desired $X_\theta$ gate into $4 X_{\theta/4}$ gates. We choose $\thetafull \in \{0,\frac{\pi}{5},\frac{2\pi}{5}...,\frac{15\pi}{5}\}$. This corresponds to a maximum $\thetagate$ of $\frac{3\pi}{4}$. In each job sent to the device for execution, we include all the circuits corresponding to the pseudoidentity $U_{\thetafull}$ and $U_0$. We find that with $\Delta n = 10$ we do not see aliasing effects, and hence use this value in our runs. With this $\Delta n$, in each job we run the circuits up to $n=150$, so that $n \in \{0,10,20,...150\}$. We use $1024$ shots to estimate the observables.

Our experiments were run between the dates 6\textsuperscript{th} May 2022 and 23\textsuperscript{rd} May 2022. Each day, we ran experiments corresponding to all $16$ values of $\thetafull$.

\subsection{Purity oscillations as signature of non-Markovianity for undriven and driven qubits}
\label{subsec:purity_oscillations_signature_non_markovian}

\begin{figure}[t!]
  \centering
  \includegraphics[width=\columnwidth]{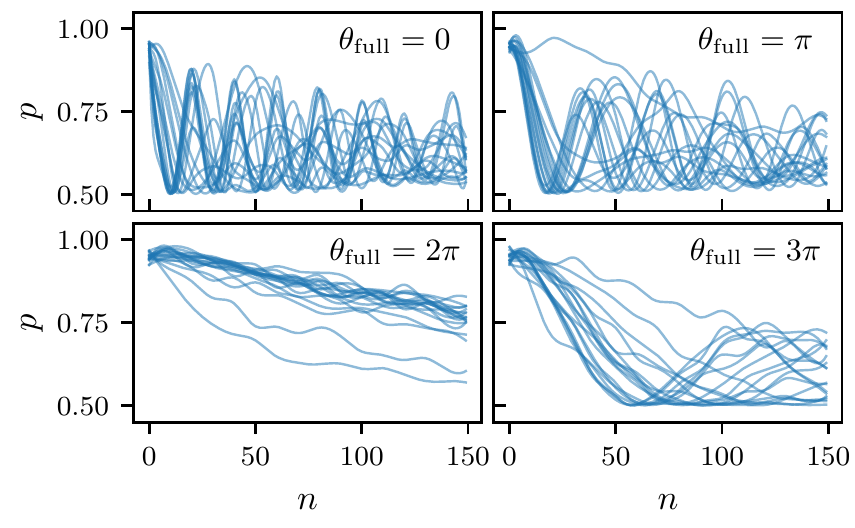}
  \caption{Purity, $p$, as function of number of applied pseudoidentities, $n$. Different curves in each graph correspond to the results on different days. The continuous trajectories are obtained from the discrete points by a cubic spline interpolation. Oscillations in the purity of the qubit are visible and the oscillation frequency can be seen to change with $\thetafull$.}
  \label{fig:purity_oscillation}
\end{figure}  
\begin{figure}[t!]
  \centering
  \includegraphics[width=\columnwidth]{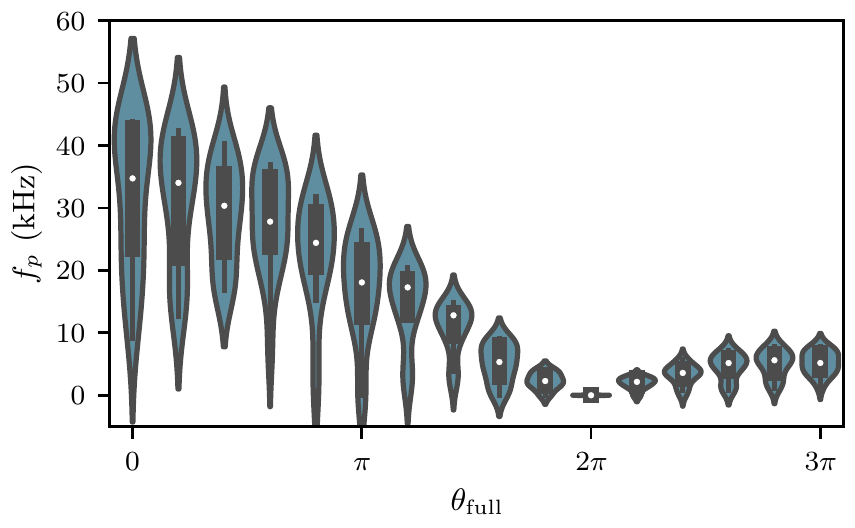}
  \caption{Distributions of the qubit purity oscillation frequencies $f_p$ obtained by fitting the purity trajectories in \figref{fig:purity_oscillation} to an exponentially decaying oscillation of form
  given in Eq. \ref{Eq:purity_oscillations}. The violin plots for each value of $\thetafull$ are obtained by kernel density estimation. The purity oscillation frequency decreases until it reaches $0$ at $\thetafull = 2\pi$ and then starts increasing again.}
  \label{fig:purity_oscillation_trend}
\end{figure}  
In \figref{fig:purity_oscillation} we plot the purity $p$ of the qubit as a function of the number of applied pseudoidentity gates $n$ applied in the experiments, for a representative set of qubit drive angles $\thetafull = \{ 0, \pi, 2\pi, 3\pi\}$. The purity is obtained with the formula  $p = \Tr{\hat{\rho}^2} = (1+\Tr{\hat{\rho} \sx}^2 + \Tr{\hat{\rho} \sy}^2 +  \Tr{\hat{\rho} \sz}^2)/2 = (1+\expval{\sx}^2 + \expval{\sy}^2 +  \expval{\sz}^2)/2 $ [\secref{subsec:nonmarkovianity}]. 
Different curves on the same plot denote results from runs over 18 different days.
Importantly, oscillations in the qubit purity can clearly be observed for $\thetafull = \{ 0, \pi, 3\pi\}$, indicating the presence of non-Markovian noise.

In \figref{fig:purity_oscillation} one can observe that the purity takes the form of a single frequency oscillation with an amplitude that decays with increasing time. Motivated by the observed shape of the curves in \figref{fig:purity_oscillation}, in order to analyse the trend in purity oscillation frequency, we fit the results to an exponentially decaying oscillation of the form
\begin{equation}
p = \frac{1 + \cos(2\pi f_p n T)^2e^{-\gamma_p n T} }{2}, 
\label{Eq:purity_oscillations}
\end{equation}
where $T$ is the duration of a single pseudoidentity operation and $f_p, \gamma_p$ denote the frequency and decay rate of the purity oscillations. The form of this equation is a special case of the general equation for non-Markovian purity oscillations given in \eqref{eq:pt_decay}. The distributions of the frequency $f_p$ for different values of $\thetafull$ are shown in \figref{fig:purity_oscillation_trend}. One can see that the frequencies of the purity oscillations reduce until $\thetafull = 2\pi$, and then start increasing again. As discussed in \secref{subsubsec:noise_characterisation_circuits}, $\thetafull = 2\pi$ is the special case where small, time-independent errors in the qubit Hamiltonian are not amplified. The absence of purity oscillations as signature of non-Markovian noise indicates  cancellations of the non-Markovian contributions to the time evolution at $\thetafull = 2\pi$. Importantly, for all other angles $f_p$ is non-zero, indicating non-Markovian dynamics in the qubit, both in its idle state and in the driven case.

\subsection{Noise model results for the undriven qubit}
\label{subsec:undriven_qubit_results}

\subsubsection{Noise model comparison for the undriven qubit}

\begin{figure}
  \includegraphics{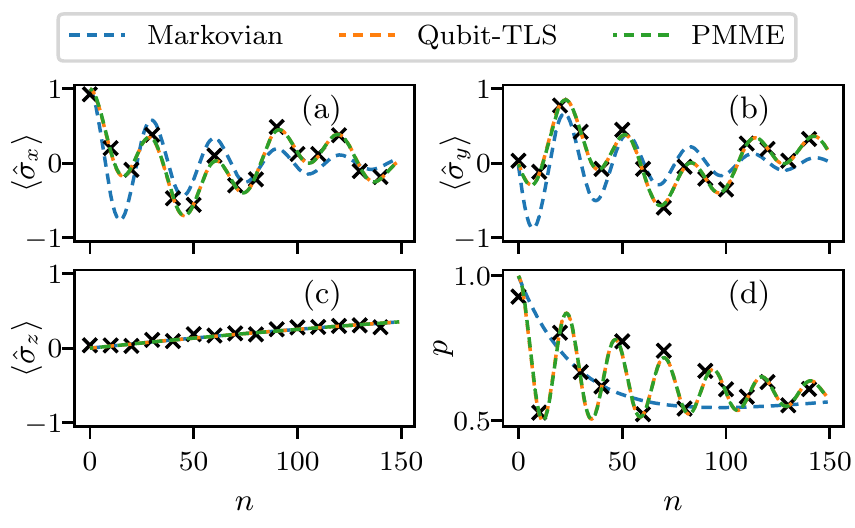}
  \caption{Example of the undriven qubit model-fitting result for experiments run on  a single day. The expectation values of the qubit in the $X,Y$, and $Z$ basis, and the qubit purity $p$ are plotted in the different subplots as a function of the number of applied pseudoidentity idling operations $n$. On each subplot, the results for all three noise models are plotted in different colours. The black crosses denote the experimental results. The non-linear regression RMSEs for the Markovian, qubit-TLS, and PMME models are 0.215, 0.044, and 0.044 respectively. The non-Markovian qubit-TLS and PMME model fit the experimental data well, while the Markovian model does not.}
  \label{fig:nm_undriven_qubit_fitting_example}
\end{figure}
\begin{figure}
  \includegraphics[width=\columnwidth]{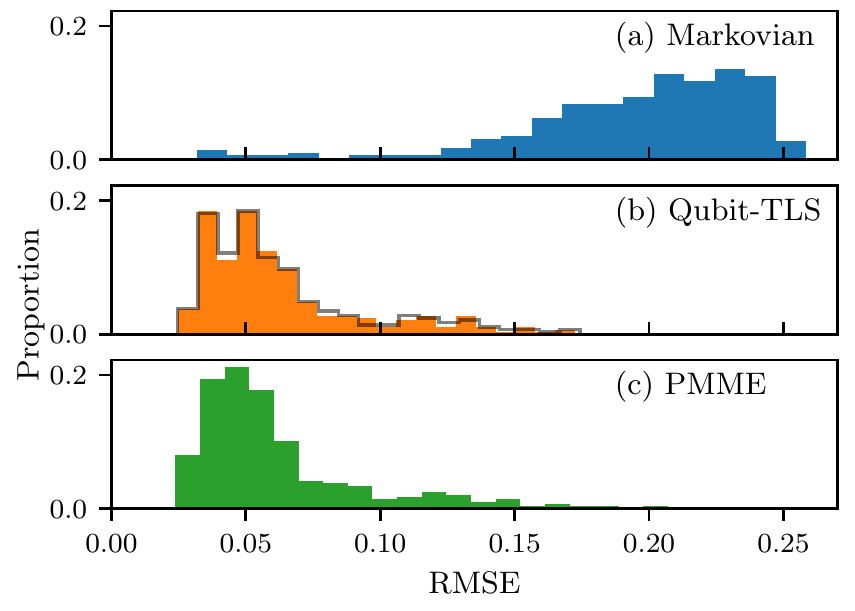}
  \caption{Distribution of the root-mean-squared-errors (RMSEs) of the different noise models for the undriven qubit. The Markovian model has much larger RMSE, corresponding to worse quality of fitting to experimental data, while the qubit-TLS model and the PMME both have good quality of fits. The solid line in (b) shows the distribution of RMSE for the qubit-TLS model in the absence of any amplitude damping on the TLS, i.e. $\kappa = 0$.}
  \label{fig:rmse_distribution_undriven_qubit}
\end{figure}
\begin{figure}
  \includegraphics[width=246pt]{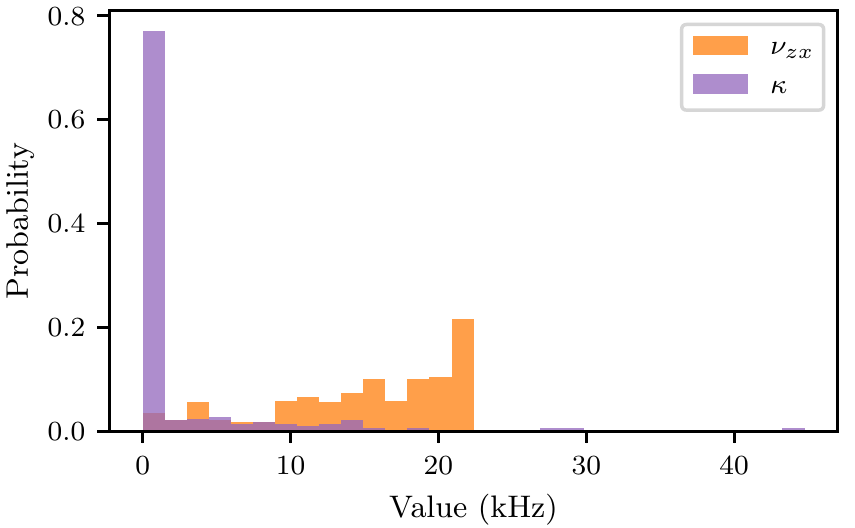}
  \caption{Distribution of the parameters $\kappa$ and $\nu_{zx}$ in the qubit-TLS model. The values of $\kappa$ can be seen to be significantly less than the values of $\nu_{zx}$ with the exception of a few outliers.}
  \label{fig:tls_amplitude_damping_distribution}
\end{figure}

Having seen that a signature of non-Markovian behaviour in the form of damped purity oscillations is present in the qubit dynamics, we now investigate how well the noise models are able to match the observed experimental results for the undriven qubit.

In \figref{fig:nm_undriven_qubit_fitting_example} we present a representative example of how the different noise models fit the experimental results obtained for the undriven qubit ($\thetafull=0$). In the experimentally obtained qubit expectation values for $\expval{\sx}$ and $ \expval{\sy}$, we find multi-frequency oscillations, and in the purity of the qubit we see damped oscillations. These features of non-Markovianity cannot be reproduced by the Markovian model, which is therefore unable to fit the experimental data. Note that this will be the case even if an arbitrarily large number of parameters are included in a general Markovian model. 

To analyse the quality of fits over all the performed idle qubit runs we plot the distribution of the RMSEs for all these runs, and for the different noise models, in \figref{fig:rmse_distribution_undriven_qubit}. We observe that the Markovian model has significantly higher RMSE than the non-Markovian models, confirming its inherent inability to fit non-Markovian dynamics. The qubit-TLS model and the PMME model, on the other hand, both have low RMSEs. We attribute the remaining lower RMSE boundary of RMSE $> 0.025$ to experimental factors that are not included in the noise model, such as SPAM errors and finite sampling noise (shot noise) [see also \secref{subsec:noise_models}]. We can estimate an upper bound on the expected shot noise by considering the maximal uncertainty case, where the two outcomes of a qubit measurement are equally likely. Using the formula for standard deviation of a binomial distribution $\sigma = \sqrt{\frac{p*(1-p)}{n}}$, where $p$ is the probability of a particular outcome and $n$ is the number of trials, we can estimate the expected shot noise as $\sigma_\mathrm{shot} = 2*\sqrt{\frac{0.5*0.5}{1024}} \simeq 0.03$. The factor of $2$ arises from the conversion of a probability, which has range $[0,1]$, to an expectation value, which has range $[-1,1]$. The expected upper bound on the shot noise of $\sim 0.03$ is approximately in agreement with the obtained lower RMSE boundary of RMSE $> 0.025$. 

In order to analyse the magnitude of the dissipation on the TLS and its contribution to the dynamics, in \figref{fig:tls_amplitude_damping_distribution} we plot the distribution of the parameters $\kappa$ and $\nu_{zx}$ in the qubit-TLS model. One can see that the vast majority of the values of $\kappa$ are very close to zero on this scale, and significantly smaller than the values of $\nu_{zx}$. This shows that the qubit-TLS coupling for this system is much stronger than the dissipation term on the TLS. In order to evaluate whether the presence of such a weak dissipation in the TLS affects the quality of fits of the qubit-TLS model results, we analyse how the RMSE distribution of the qubit-TLS model changes when the value of $\kappa$ is fixed to be $0$. In \figref{fig:rmse_distribution_undriven_qubit} (b), the solid black line denotes the RMSE distribution for the qubit-TLS model in the absence of amplitude damping on the TLS, i.e. $\kappa = 0$. We can see that there is no noticeable change in the RMSE distribution when we apply this restriction. This shows that a model which has the qubit interacting with a non-dissipative TLS can fit the experimental results equally well. Thus, in the remainder of this article, we use a qubit-TLS model without dissipative term on the TLS, corresponding to $\kappa = 0$.

The qubit-TLS and PMME RMSE distributions in \figref{fig:rmse_distribution_undriven_qubit} are almost identical. This is due to the fact that the two models are mathematically equivalent to each other, as discussed in \secref{subsec:PMME_qtls_relation}. The near-identical distributions for the two models validate the non-linear regression process in both cases, since they both find the appropriate and equivalent minima completely independently from each other. The fact that one can get an equally good fit fixing $\kappa=0$ in the qubit-TLS model shows that it possesses an excess parameter, $\kappa$, beyond what is required to accurately fit the experimental results for this device. Given the equivalence of the PMME and qubit-TLS model, this also implies that the PMME model has one excess parameter. Both models can therefore be considered over-parameterized within the context of the performed experiments. We discuss the over-parameterisation in the qubit-TLS and PMME model further in \appref{app:overparameterisation}, and we numerically verify the relations between the qubit-TLS and PMME model parameters in \appref{app:pmme_qtls_relation}.

The PMME cannot directly be used when the Hamiltonian has a time-dependence, which is the case for the driven qubit. Thus, in this article we limit our analysis of the PMME model to the undriven qubit, although an extension to the driven qubit can be derived, and will be presented as part of future work. 

\subsubsection{Qubit-TLS model noise parameter fluctuations}
\begin{figure*}
    \centering
    \includegraphics[width=\textwidth]{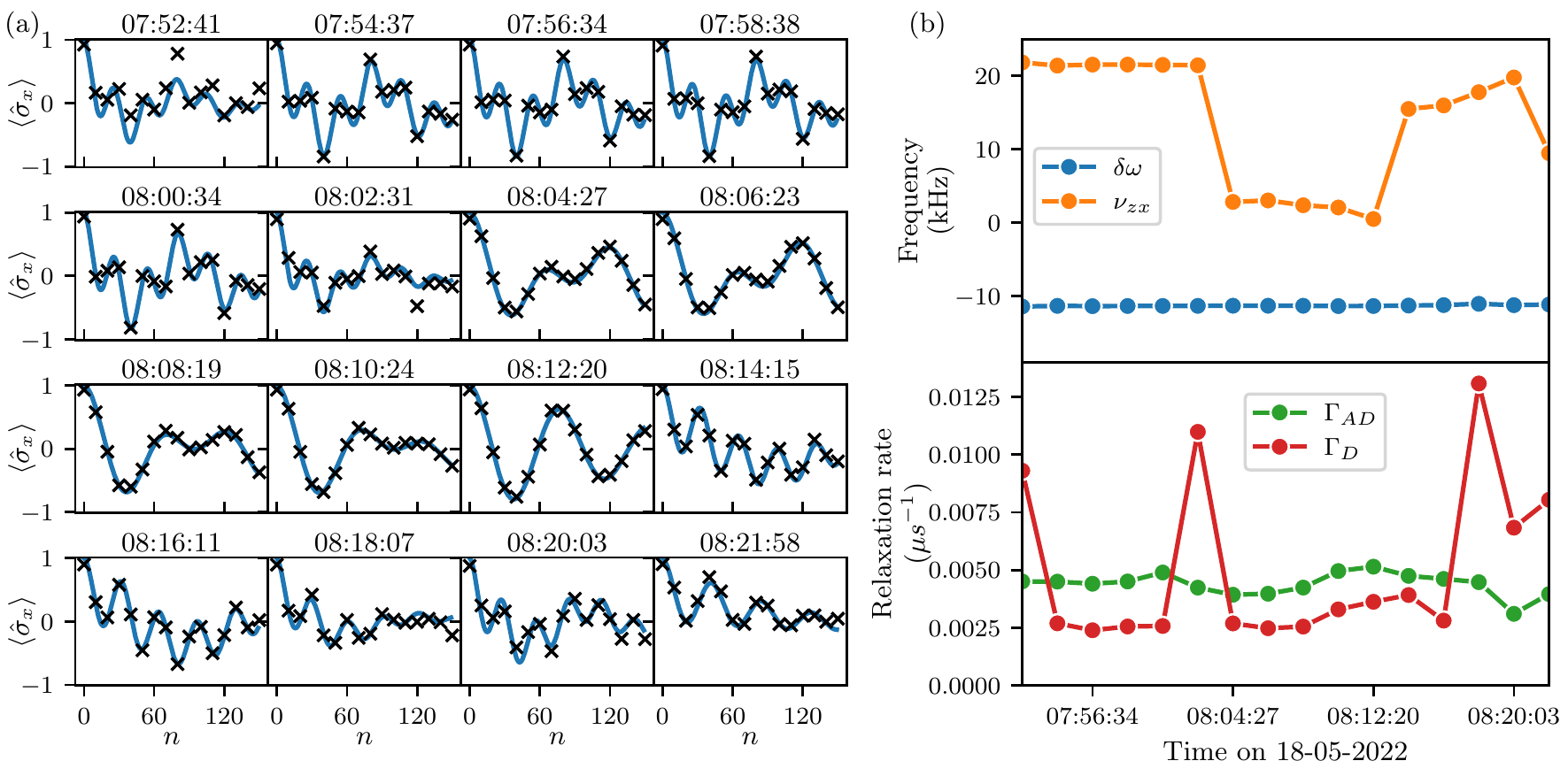}
    \caption{Changing behaviour of the idle qubit evolution on circuits batches run consecutively on a single day. (a) shows the expectation value $\expval{\sx}$ for $\theta = 0$ as a function of $n$ at different times while (b) shows the corresponding qubit-TLS noise model parameters as a function of time. Sudden, large jumps in the noise model parameters are visible.}
 \label{fig:single_day_drift_example}
\end{figure*}

We now investigate how the observed dynamics of the idle qubit changes with time, and how it affects the obtained noise model parameters. As a representative example we show the results for the idle qubit obtained on 18\textsuperscript{th} May 2022 in \figref{fig:single_day_drift_example}. Each subplot corresponds to results obtained from circuits sent in different batches. The 16 ``jobs" were sent consecutively to the device and completed in the span of approximately 30 minutes.

In \figref{fig:single_day_drift_example} (a), we can see that the qubit dynamics can change significantly even between consecutive jobs run within $2$ minutes. This is most clearly visible for the sudden change between the dynamics at 8:02:31 \& 8:04:27, and at 08:12:20 \& 08:14:15.

In order to evaluate which noise parameters determine the found sudden changes, we analyse the noise model parameters for all the presented times in \figref{fig:single_day_drift_example} (a). The results are shown in \figref{fig:single_day_drift_example} (b). One can see that during these sudden changes in dynamics, the parameter corresponding to the qubit-TLS interaction, $\nu_{zx}$, changes significantly. This shows that the qubit dynamics can be very unstable over time due to the non-Markovian noise changing abruptly.
The big jumps in $\nu_{zx}$ can have various physical origins. For example, the coupled TLS can shift in energy due to interaction with other TLS defects, which have low-frequency switching rates \cite{burnettDecoherenceBenchmarkingSuperconducting2019,klimovFluctuationsEnergyRelaxationTimes2018,burnettEvidenceInteractingTwolevel2014,mullerInteractingTwolevelDefects2015} ,or due to background ionizing radiation \cite{thorbeck2022tls}.

The parameters $\delta\omega$ and $\Gamma_{AD}$, on the other hand, do not change significantly in these $30$ minutes. The parameter $\Gamma_{D}$ does jump significantly at some specific times; however, one can see that the high values of $\Gamma_D$ occur when the corresponding fit is less accurate than the other fits. This is most clearly visible in the results at 07:52:41 and 08:02:31. If the results at these times were significantly affected by sources of noise not considered in the noise model, the model will be incapable of accurately fitting the experimental data. This can then lead to the resulting dephasing noise parameter not corresponding to the a physical value but rather corresponding to a value that minimises the RMSE subject to the limitations of the noise model. 

\begin{figure}
  \includegraphics[width=\columnwidth]{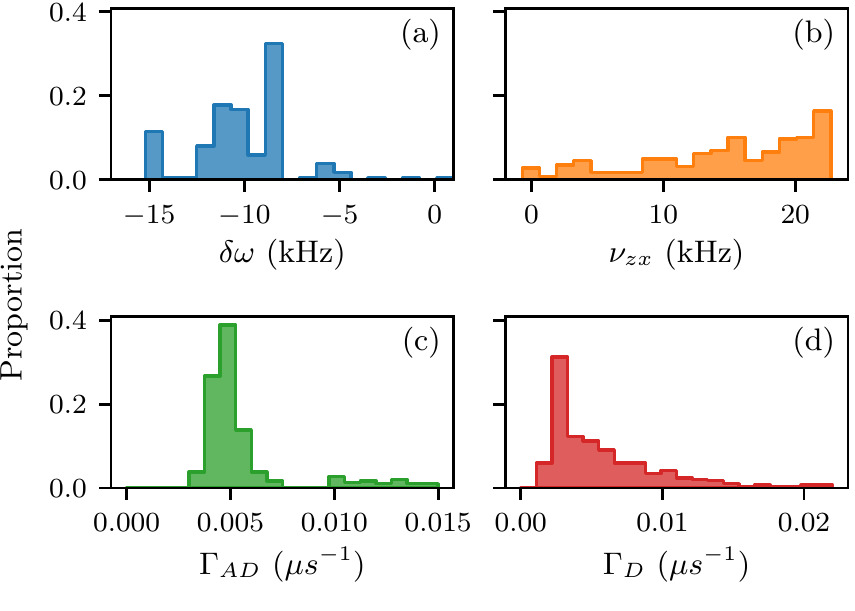}
  \caption{Distribution of the qubit-TLS noise model parameters for the idle qubit, i.e. $\thetafull = 0$.}
  \label{fig:nm_3_undriven_noise_parameter_distribution}
\end{figure}
Having seen that the experimental results can change significantly even within the time-span of a few minutes due to the sudden change in the non-Markovian noise, we now analyse the distribution of the noise model parameters over the 18 days in \figref{fig:nm_3_undriven_noise_parameter_distribution}. We can see that although the $\delta\omega$ and $\Gamma_{AD}$ parameters seemed very stable over the course of the $30$ minutes in \figref{fig:single_day_drift_example}, which would lead to a very narrow distribution, both the parameters have a wider distribution when the data is aggregated over multiple days.

In order to analyse the time-scales of the fluctuations of each parameter, in \tabref{tab:std_timescale} we show the overall standard deviation of the noise parameters obtained over the $18$ days, together with the mean of the standard deviations for results obtained on the same day. For both $\delta\omega$ and $\Gamma_{AD}$ we can see that the mean of the daily standard deviation is much lower than the overall standard deviation. On the other hand, for $\nu_{zx}$ and $\Gamma_D$ the two values are similar. This indicates that the parameters $\delta\omega$ and $\Gamma_{AD}$ do not fluctuate significantly in the time scales of $\sim 30$ minutes but they can fluctuate significantly in the time scale of a few days or weeks. Conversely, the parameters $\nu_{zx}$ and $\Gamma_D$ can fluctuate significantly already in the time scales of a few minutes. However, as noted before in the discussion of \figref{fig:single_day_drift_example}, inaccuracy in some of the fits can lead to jumps in the dephasing noise parameter. Thus, rapid fluctuations of $\Gamma_{D}$ can be an artifact of the limitations of the noise model.

\begin{table}[ht]
  \centering
  \begin{tabular}{|c|c|c|}
    \hline
    \begin{tabular}{@{}c@{}}\textbf{Noise} \\ \textbf{parameter} \end{tabular}  &  \begin{tabular}{@{}c@{}}\textbf{Overall} \\ \textbf{standard deviation} \end{tabular} & \begin{tabular}{@{}c@{}}\textbf{Mean of daily} \\ \textbf{standard deviation} \end{tabular}  \\
    \hline
    $\delta\omega (\unit{\kilo\hertz})$ & $2.3$ & $0.12 \pm 0.06$ \\
    $\nu_{zx} (\unit{\kilo\hertz})$ & $6.2$ & $5.7 \pm 1.5$ \\
    $\Gamma_{AD} (\unit{\per\micro\second})$ & $0.0025$ & $0.0005 \pm 0.0002$ \\
    $\Gamma_{D} (\unit{\per\micro\second})$ & $0.0038$ & $0.0027 \pm 0.0014$ \\
    \hline
  \end{tabular}
  \caption{Comparison of overall standard deviation of the noise parameters with the mean of the standard deviations for results obtained on a single day. Note that we have only taken into account the fitting results which had RMSE $< 0.1$ in order to minimize the effects of the low fitting quality results.}
  \label{tab:std_timescale}
\end{table}

\subsection{Noise model results for the driven qubit}
\label{subsec:driven_qubit_results}

\subsubsection{Noise model comparison for the driven qubit}
\begin{figure*}
  \includegraphics[width=\textwidth]{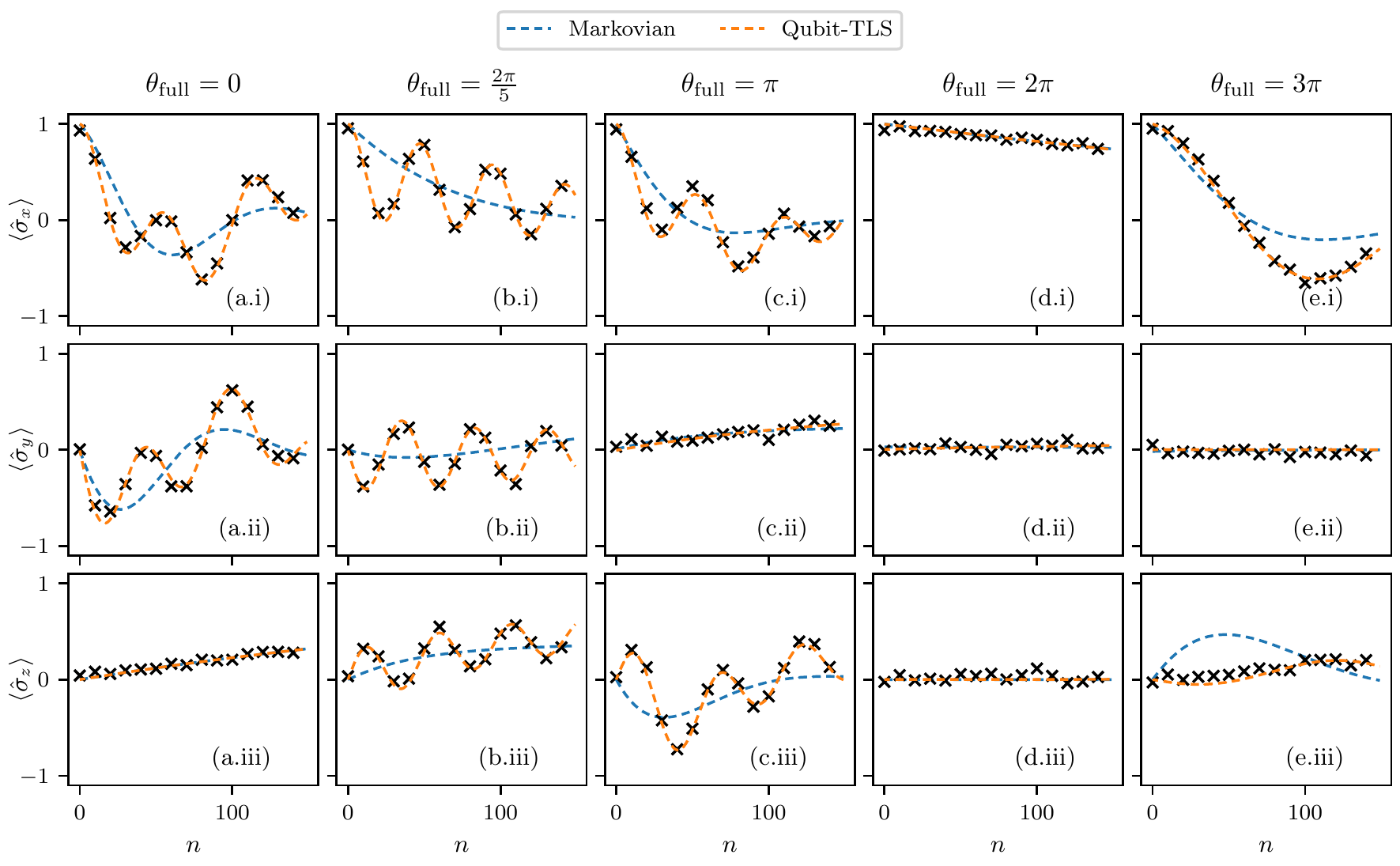}
  \caption{Examples of the non-linear regression results for experiments run on 17\textsuperscript{th} May 2022. Different rows represent different amounts of rotation $\thetafull$ applied during the driving. The columns correspond to measurements in the $X,Y$, and $Z$ basis respectively. The results for the different noise models are shown in different colours. The black crosses denote the experimental results. }
  \label{fig:fitting_example}
\end{figure*}
\begin{figure}
  \includegraphics[width=\columnwidth]{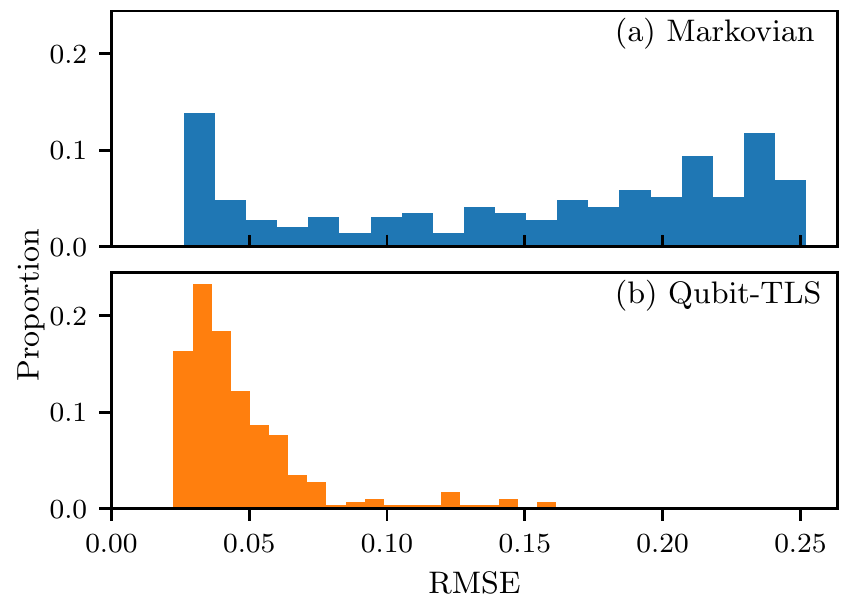}
  \caption{Distribution of the root-mean-squared-errors of the different noise models for all values $\thetafull$ and all $18$ days. The non-Markovian qubit-TLS model has significantly smaller RMSE than the Markovian model.}
  \label{fig:rmse_distribution_driven_qubit}
\end{figure}

In \figref{fig:fitting_example} we show a representative set of experimental results that we obtain on a single day for a range of $\thetafull$, together with the fitted noise model curves.
The undriven qubit case results, corresponding to $\thetafull = 0$, show again that the Markovian model is not able to describe the experiment, while the non-Markovian qubit-TLS model is. The only $\thetafull$ for which the Markovian model works well is $\thetafull = 2\pi$. As discussed in \secref{subsubsec:noise_characterisation_circuits} and \appref{app:hamiltonian_error_insensitivity}, this is the special case for which the qubit dynamics is insensitive to small time-independent errors in the qubit Hamiltonian. In \ref{fig:fitting_example} (d), we do not see any oscillations of the observables, which is consistent with the absence of purity oscillations for $\thetafull=2\pi$ discussed in \secref{subsec:purity_oscillations_signature_non_markovian}. Thus, we can see that the non-Markovian noise is effectively cancelled out by this gate sequence at $\thetafull=2\pi$, which allows the Markovian model to fit the experimental data at this angle.

\begin{figure}
  \includegraphics[width=\columnwidth]{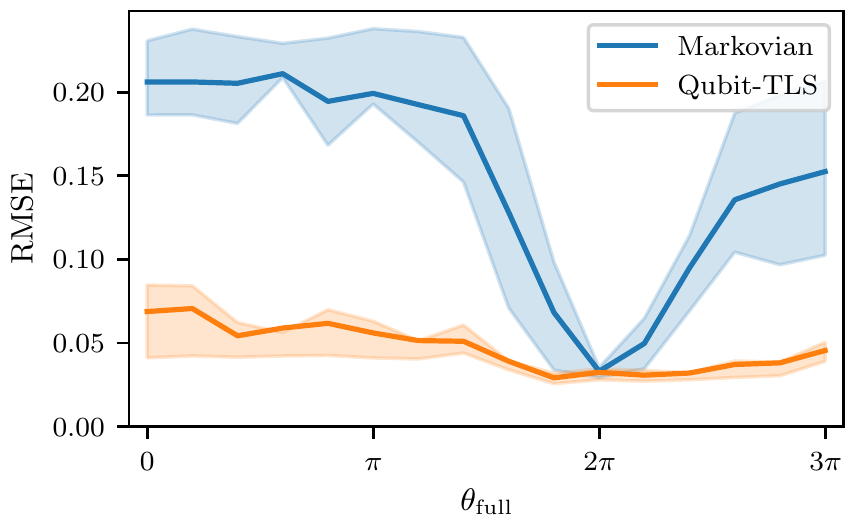}
  \caption{Mean of the root-mean-squared-error for the different noise models as a function of $\thetafull$. The shaded region denotes the region between the upper and lower quartiles. The quality of fits of the qubit-TLS model doesn't change significantly, while the Markovian model only has a low RMSE near the special case $\thetafull = 2\pi$.}
  \label{fig:mse_vs_theta}
\end{figure}

The qubit-TLS model fits the experimental data well for all the $\thetafull$. In order to see if this holds true across all the acquired data for all days and times, we plot the distribution of the RMSEs for both noise models in \figref{fig:rmse_distribution_driven_qubit}, and we plot their RMSE as a function of $\thetafull$ in \figref{fig:mse_vs_theta}.

In \figref{fig:rmse_distribution_driven_qubit} one can see that in the qubit-TLS model, the distribution of the RMSEs has a pronounced peak at $\sim 0.025$, corresponding to good quality fits, in analogy to what we found for the undriven qubit in \figref{fig:rmse_distribution_undriven_qubit}. The Markovian model on the other hand shows a much flatter distribution of the RMSEs up to high errors in the fit, also in agreement with the undriven qubit results. The $\thetafull$ dependence of the quality of fits presented in \figref{fig:mse_vs_theta} shows that the qubit-TLS model works very well for the entire $\thetafull$ range, while the Markovian model only fits the data well near $\thetafull = 2\pi$, which is the case where there are no signatures of non-Markovian behaviour in the experimental data, as discussed above. The fact that the qubit-TLS model can capture the experimental behaviour well over a large number of days and angles shows that it captures the physical process that dominate the noise, including both the Markovian and non-Markovian components.

\subsubsection{Change of qubit-TLS model parameters with drive pulse amplitude}
\label{subsec:qubit_tls_driven_qubit_analysis}

 \begin{figure*}
 
     \includegraphics[width=\textwidth]{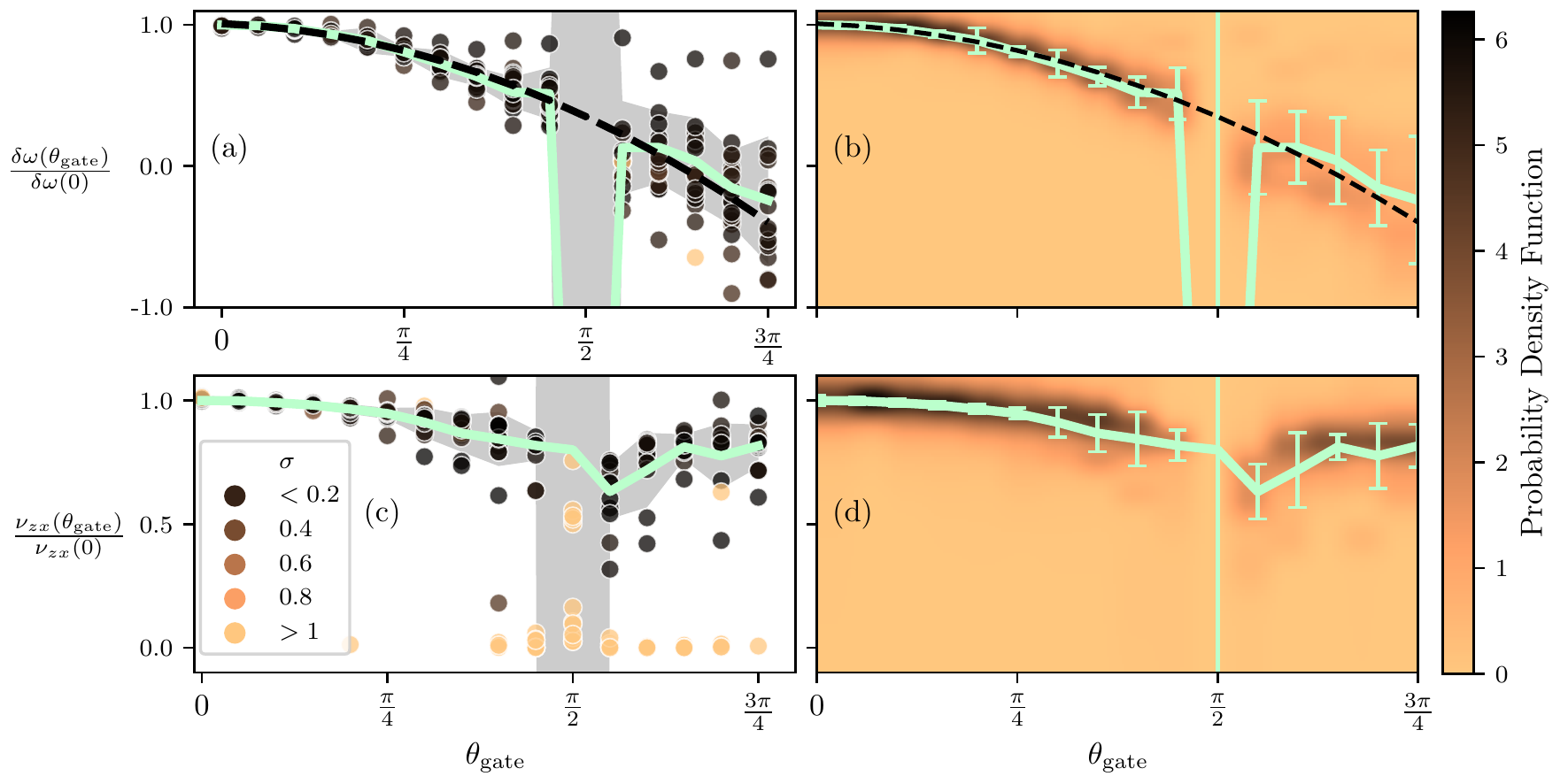}
  \caption{Weighted means of the ratio of the noise model parameters $\delta\omega$ (a,b) and $\nu_{zx}$ (c,d) for the driven and idle qubit as a function of $\thetagate$. Plots (a,c) show the measured values of the noise parameters as dots, with the colour of the dots representing the uncertainty $\sigma$ due to the non-linear regression. The green line and shaded region in (a,c) denote the weighted mean and total uncertainty in the weighted mean respectively. Plots (b,d) show the combined distribution of the dots in (a,c) obtained by converting each measurement into a Gaussian distribution centred at the measured value with width $\sigma$. The ratio of $\delta\omega$ can be seen to vary approximately quadratically with $\thetagate$. The dashed lines in (a,b) show the quadratic curve the best fits the weighted mean. The ratio of $\nu_{zx}$ shows a non-monotonic behaviour with $\thetagate$ but does not change as significantly as $\delta\omega$. The large uncertainties at $\thetagate = \pi/2\implies\thetafull = 2\pi$ are due to the insensitivity of the qubit dynamics to constant errors in the qubit Hamiltonian.}
  \label{fig:nm_3_driven_parameter_ratio}
 \end{figure*}
We now investigate the relative changes of the qubit-TLS noise model parameters for the different values of $\thetagate=\thetafull/4$. As discussed in \secref{subsubsec:noise_characterisation_circuits} and \secref{subsubsec:non_linear_regression_data_analysis}, due to the large variation in qubit behaviour observed over the time scale of a few minutes, one cannot directly compare noise model parameters corresponding to circuits that have been run in different batches of jobs. Since each batch of circuits that we submit includes the circuits corresponding to the idle qubit and to a qubit driven with pseudoidentities with a corresponding angle $\thetafull$, the ratio of the noise parameters for the driven and idle qubit can act as a metric that we can use to analyse how the noise parameters change with $\thetafull$ and hence with $\thetagate$, since $\thetafull = 4 \thetagate$. Note that it is possible that even within one batch of circuits there is a sudden jump in the experimental noise properties, which can lead to some outliers in our data. We find that such outliers constitute only a very small fraction of the observed results, and that they also have a large fitting uncertainty.

To obtain the ratio of the noise parameters, we use the noise model fitting procedure described in \secref{subsubsec:non_linear_regression_data_analysis}, which involves combined regression with the results corresponding to a $\thetagate$ and the undriven qubit. As discussed in \secref{subsubsec:non_linear_regression_data_analysis}, having excess parameters in the noise model can lead to over-fitting. It is thus beneficial to minimize the number of free parameters in the model. In the combined regression, we can reduce the total number of parameters involved in the fitting by constraining some of the parameters to be the same for the driven and undriven qubit.
In order to evaluate which of the noise model parameters can be constrained in this manner, we analyse the change in the RMSE distribution when individual parameters were constrained to be the same for the driven and undriven qubit. We find that RMSEs increase significantly when we set $\delta\omega(\thetagate) = \delta\omega(0)$ or $\nu_{zx}(\thetagate) = \nu_{zx}(0)$, but the RMSEs do not change significantly when we set $\Gamma_D(\thetagate) = \Gamma_D(0)$ or $\Gamma_{AD}(\thetagate) = \Gamma_{AD}(0)$. Constraining both the amplitude damping and dephasing noise parameters simultaneously only marginally reduces the overall accuracy of the non-linear regression: the average mean-squared-error increases by $17\%$. Therefore, in the results presented in this section, we constrain the dephasing and amplitude damping parameters in order to minimize over-fitting.

We now describe an analogue to kernel density estimation \cite{Parzenkde} that can be used to visualise the distribution of the noise parameters. Let $r_{ik}(\thetafull)$ and $\sigma_{ik}(\thetafull)$ denote a noise parameter ratio and its corresponding uncertainty with the noise kind indexed by $i$, and the date of the experiment indexed by $k$. To visualise the distribution of the $r_{ik}(\thetafull)$ while also taking into account the sources of uncertainties discussed in \secref{subsubsec:non_linear_regression_data_analysis}, we replace each value of $r_{ik}(\thetafull)$ with a Gaussian distribution with mean $r_{ik}(\thetafull)$ and standard deviation $\sigma_{ik}(\thetafull)$. Let $f_{ik}(\thetafull,z)$ correspond to the probability density function of a random variable $z$ corresponding to these Gaussians. Then we can combine the distributions by calculating
\begin{equation}
\tilde{f}_{i}(\thetafull, z) = \frac{1}{K} \sum_k f_{ik}(\thetafull,z).
\end{equation}
This probability density function can be plotted for each $\thetafull$ to obtain a visual representation of how the noise model parameter $x_i(\thetagate=\thetafull/4)$ changes with $\thetagate$.

In \figref{fig:nm_3_driven_parameter_ratio}, we plot the weighted mean of the ratio of the noise model parameters $\delta\omega$ and $\nu_{zx}$ for the driven and idle qubit as a function of $\thetagate=\thetafull/4$. The weighted mean, with the weights being given by $1/\sigma_r^2$, where $\sigma_r$ is the non-linear regression uncertainty in the ratio of the noise parameters, de-weights the results with large uncertainties, making it more robust against high-uncertainty outliers than the standard mean. The method to obtain the total uncertainty associated with the weighted mean is discussed in \secref{subsubsec:non_linear_regression_data_analysis}. We show the weighted means and uncertainties of both noise parameters in two different representations: in \figref{fig:nm_3_driven_parameter_ratio} (a,c) we show calculated noise parameter ratios as dots, with their color representing the fitting uncertainty; in \figref{fig:nm_3_driven_parameter_ratio} (b,d) plot the probability density function of the noise model parameters using the representation presented above. 
These combined Gaussian distributions, shown in colour in \figref{fig:nm_3_driven_parameter_ratio} (b,d), firstly indicate that there is a narrow spread in the values of the obtained ratios even for $\thetagate = 0$. This indicates the uncertainty in the calculated ratio for the undriven qubit. This uncertainty includes the effects of finite sampling error, as well as errors due to small changes in the noise parameters that can take place within a single batch of circuits. Importantly, although the values of the individual parameters themselves change significantly over the different days, the relative change of parameters with drive amplitude has a well defined behaviour across all the days.

For $\thetagate = \pi/2$, the ratios have a very large uncertainty. This is expected because the dynamics is insensitive to errors such as the phase error at $\thetafull = 4\thetagate= 2\pi$, as discussed in \secref{subsubsec:noise_characterisation_circuits}.  
In  \figref{fig:nm_3_driven_parameter_ratio} (a,b) one can see that $\frac{\delta \omega(\thetagate)}{\delta \omega(0)}$ changes quadratically with $\thetagate$. The amount of change is large: the ratio goes from $1$ to $\sim -0.25$ within the $\thetagate$ range that we consider. This quadratic behaviour can be due to the AC Stark effect causing a frequency shift of the qubit, which depends on the power of the applied pulses \cite{schneiderLocalSensingMultiLevel2018}.

In  \figref{fig:nm_3_driven_parameter_ratio} (c,d) we see that the value of $\frac{\nu_{zx}(\thetagate)}{\nu_{zx}(0)}$ shows a non-monotonous behaviour with $\thetagate$. The ratio becomes smaller as $\thetagate$ is increased from $0$ to $\pi/2$, however, it does not reduce as significantly as the ratio of $\delta \omega$. The large uncertainties near $\thetagate = \pi/2$ make it difficult to evaluate how $\nu_{zx}$ is changing near those values. For $\thetagate>\pi/2$, the $\nu_{zx}$ ratio seems to increase with $\thetagate$ before plateauing. Note that even the $20\%$ reduction in $\nu_{zx}$ from $\thetagate = 0$ to $\thetagate = 3\pi/4$ is significant; if we constrain $\nu_{zx}(\thetagate)$ to be equal to $\nu_{zx}(0)$, as we do for the amplitude damping and dephasing noise parameters, the performance of the nonlinear regression is significantly worse.

We note here the importance of quantifying uncertainties in the non-linear regression. If we would aggregate the results over different days by performing a standard mean, high-uncertainty outliers, such as the $\frac{\nu_{zx}(\thetagate)}{\nu_{zx}(0)}\sim 0$ values seen in \figref{fig:nm_3_driven_parameter_ratio} (c), would significantly affect the calculated mean. Taking into account the fitting uncertainties in those measurements allows us to de-weigh the effects of such high-uncertainty outliers.

We note that we performed the fit for noise models with additional contributions as well. These included models with the $\epsilon_z\hat{\tau}_z$ or $\nu_{zz}\sz \otimes\hat{\tau}_z $ terms included in the Hamiltonian. These lead to only slightly improved overall accuracy. The main drawback of using such over-parameterised models is that parameters such as $\nu_{zx}$ do not vary smoothly with $\thetagate$ anymore, and instead show a wide and erratic distribution. This is a sign of over-fitting, and the trends extracted from such over-fitted cannot be trusted. By requiring noise models to have parameters which change smoothly with the driving strength, we can avoid  noise models that over-fit the experimental data. Thus, considering the dynamics of driven qubits not only allows us determine how the noise parameters change with the driving strength, but it can also be used to guide the development of a noise model specific to specific hardware.

\section{Summary and Conclusions}
We analyze non-Markovian noise present in superconducting qubits by developing and running gate sequences involving mirrored-pseudoidentity operations. In the experimental results, we find signatures of non-Markovianity, including damped oscillations of qubit purity and multi-frequency oscillations of qubit observables as a function of the number of applied pseudoidentity gates. 
We compare a Markovian noise model, a noise model introducing non-Markovianity via the post Markovian master equation, and a noise model including non-Markovianity through an interaction of the qubit with a TLS. For the idle qubit, we find that the PMME model and the qubit-TLS model can both successfully fit the experimental data very well, while the Markovian model does not fit the experimental data well and fails to reproduce the observed signatures of non-Markovianity. Analysis of the qubit-TLS model results reveals that a non-dissipative TLS is sufficient to model the obtained experimental results.
By analysing the results obtained at different times, we find that the qubit dynamics can fluctuate significantly even on the time scales of a few minutes. Using the qubit-TLS results, we find that this is caused by the instability of the non-Markovian noise. We also find that the the phase error and amplitude damping noise parameters in the model fluctuate over time, albeit at slower time scales. 

We analyze the effects of non-Markovianity when the qubit is driven by pulses of varying strengths. We fit the Markovian and qubit-TLS noise models to the experimental results, and observe that the qubit-TLS model is able to fit the experimental results very well even in the presence of driving, while the Markovian model fails to do so. We show that the phase error on the qubit and the qubit-TLS interaction strength change significantly with the amplitude of the applied pulse. Although the noise parameters themselves fluctuate significantly over different days, the relative changes in the phase error and in the qubit-TLS interaction exhibit well defined trends as functions of the pulse amplitude, within the computed uncertainties, and are consistent over all days.

The non-Markovian noise characterisation methods that we develop, along with qubit-TLS noise model, can be used to not only learn about the sources of noise in a device and how they change with time and with pulses of different amplitudes, but also to develop error mitigation and correction methods, and to guide the improvements in devices needed to reduce the effects of noise.

\section{Acknowledgements}
We thank Tobias Lindstr\"{o}m, Sebastian de Graaf, Bal\'{a}zs Gul\'{a}csi, Alexander Schnell, and Annanay Kapila for useful discussions.
We acknowledge the use of IBM Quantum services for this work. We acknowledge the support of the UK government department for Business, Energy and Industrial Strategy through the UK National Quantum Technologies Programme. The views expressed are those of the authors, and do not reflect the official policy or position of IBM or the IBM Quantum team.

\clearpage
\input{appendix.tex}

\bibliography{references}

\end{document}

%% file: appendix.tex
\appendix

\onecolumngrid

\section{Purity oscillations as a signature of non-Markovian behaviour}
\label{app:purity_oscillation_proof}
Here we show that damped oscillations of the qubit purity of the form $p(t) = \frac{1}{2} + d(t)\cos^2\qty(f_p t),
$ and $p(n) = \frac{1}{2} + d(n)\cos^2(\pi \frac{n}{2N})$, where $d(n)$ is a non-periodic function and $n,N$ are both integers, cannot be generated by a single qubit undergoing Markovian evolution with $n$ repeated applications of the dynamical map $\mathcal{E}$, defined in \secref{subsec:nonmarkovianity}, such that $\mathcal{E}_n^{n+1} = \mathcal{E}_0^{1}$ for all $n$. More specifically, we show that if the purity has the above form, $d(n)$ needs to be a periodic for Markovian dynamics.

Suppose that the purity of a single qubit undergoing a Markovian evolution is given by 
\begin{equation}
p(n) = \frac{1}{2} + d(n)\cos^2(\frac{n\pi }{2N}).
\end{equation}
This implies that
\begin{equation}
p(n=N(1+2m)) = \frac{1}{2},
\end{equation}
where $m$ is an integer.
The only density matrix with purity $\frac{1}{2}$ is $\frac{1}{2}\mathds{1}$ because $p = \frac{1}{2}$ implies $\expval{\sx} = \expval{\sy} = \expval{\sz} = 0$ and $\hat{\rho} = \frac{1}{2}\mathds{1}$ is the only density matrix that satisfies that relation. Thus,
\begin{equation}
\hat{\rho}(N(1+2m)) = \hat{\rho}(N) = \frac{\mathds{1}}{2}.
\label{eq:app_purity_oscillation_proof_1}
\end{equation}
Since a Markovian evolution- by the definition presented in \ref{subsec:purity_oscillations_signature_non_markovian} - is divisible, the density matrix can be written as
\begin{equation}
\begin{split}
\hat{\rho}(N(1+2m)+q) =& \mathcal{E}_{N(1+2m)}^{N(1+2m)+q}[\hat{\rho}(N(1+2m))].\\
\end{split}
\end{equation}
Since the dynamical map is repeated, $\mathcal{E}_{N(1+2m)}^{N(1+2m)+q} = \mathcal{E}_{N}^{N+q}$. Using \eqref{eq:app_purity_oscillation_proof_1}, we obtain
\begin{equation}
\begin{split}
\hat{\rho}(N(1+2m)+q) &= \mathcal{E}_{N}^{N+q}[\hat{\rho}(N)]= \hat{\rho}(N+q).\\
\end{split}
\end{equation}
The purity is then  
\begin{equation}
p(N(1+2m)+q) = p(N+q),
\end{equation}
which finally leads to 
\begin{equation}
d(N(1+2m)+q) = d(N+q).
\end{equation}
This is valid for all $m,q\geq0$.
Thus, Markovianity implies that $d(n)$ must be a periodic function. In the limit $N \gg 1$, we can generalise from the discrete time dynamics to the continuous time dynamics, resulting in the requirement that purity oscillations of the form
\begin{equation}
p(t) = \frac{1}{2} + d(t)\cos^2\qty(f_p t),
\end{equation}
must have a periodic $d(t)$ if the governing dynamics is Markovian. Note that the equation above is only valid at stroboscopic times, $t=nT$, where $T$ is the duration of a single pseudoidentity and $n$ is an integer corresponding to the number of pseudoidentities. 

The above result shows that $d(t)$ cannot be a decaying function such as $d(t) = e^{-\gamma t}$, where $\gamma \neq 0$, in the presence of only Markovian noise. Therefore the purity cannot exhibit damped oscillations of the form given in Eq. \ref{Eq:purity_oscillations}.
\section{Single-frequency oscillation of observables for repeated Markovian processes}
\label{app:repeated_markovian_process_proof}
We show that a repeated single-qubit Markovian process will result in an oscillation of any observable with at most $1$ frequency. A single qubit density matrix can be written in the Pauli basis, i.e., $\{ \frac{\hat{I}}{2}, \sx, \sy, \sz\}$. Then, an arbitrary trace-preserving linear map can be represented by the real matrix
\begin{equation}
G = \left(\begin{array}{cccc} 1 &0 &0 &0\\c_{1,0} &c_{1,1} &c_{1,2}&c_{1,3} \\ c_{2,0} &c_{2,1} &c_{2,2}&c_{2,3}  \\ c_{3,0} &c_{3,1} &c_{3,2}&c_{3,3} \end{array}\right),
\end{equation}
where trace preservation determines the form of the first row. The form of the result of repeated applications of this map can be simplified by performing an eigenvalue decomposition such that $G = UDU^\dagger$ where $D$ contains the eigenvalues of $G$ on the diagonal. Note that this assumes that the process $G$ is reversible. Then, $G^n = UD^nU^\dagger$. Due to the form of the first row of $G$, it always has $1$ as an eigenvalue. Since $G$ is a real matrix, any complex roots must come in pairs; thus, the eigenvalues of $G$ can have at most one pair of complex roots. Assume now that the eigenvalues do include a pair of complex roots $re^{\pm i\theta}$, as otherwise the repeated single-qubit Markovian process will only give rise to exponential decay (or growth). This is because oscillations can only be generated by pairs of complex roots. We can write $G^n$ as
\begin{equation}
G^n = U\left(\begin{array}{cccc} 1 &0 &0 &0\\0 & r^ne^{in\theta} &0&0\\ 0 &0 &r^ne^{-in\theta}&0  \\ 0 &0 &0&d^n \end{array}\right) U^\dagger,
\end{equation}
where $\theta,r,d\in \mathbb{R}$. Let $\overrightarrow{\textbf{a}}(n) = (\langle\sx\rangle(n),\langle\sz\rangle(n),\langle\sz\rangle(n))$ denote the Bloch sphere coordinates of the density matrix after $n$ applications of $G$. Then,
\begin{equation}
\langle\sx\rangle(n) = f_0 + f_1 r^n\cos(n\theta + f_2) + f_3 d^n,
\end{equation}
where the values $f_i$ are functions of $\overrightarrow{\textbf{a}}(0)$ and $U$.  Similar expressions hold for $\langle\sy\rangle(n),\langle\sz\rangle(n)$. Since the operator corresponding to any single qubit observable can be written as a linear combination of $\sx,\sy,\sz$, and the identity operator (which does not evolve under the trace-preserving linear map), the expectation value of any single qubit observable can be written as
\begin{equation}
\langle\hat{O}\rangle(n) = g_0 + g_1 r^n\cos(n\theta + g_2) + g_3 d^n,
\end{equation}
where $g_i$ are functions of $\overrightarrow{\textbf{a}}(0)$ and $U$. This clearly has at most $1$ oscillation frequency. Thus, oscillations with more than $1$ frequency are signatures of non-Markovianity.

\section{Hamiltonian error insensitivity for \texorpdfstring{$\theta = 2\pi$}{theta=2pi}}
\label{app:hamiltonian_error_insensitivity}

Here we show that for the mirrored pseudoidentity with $\theta = 2\pi$ and the initial state $|+\rangle$ , the qubit dynamics become insensitive to small, time-independent errors in the Hamiltonian of the qubit of form $\h_\mathrm{ideal}(t) \rightarrow \h_\mathrm{ideal}(t) + \epsilon\h_\mathrm{noise}$. We first show that the dynamics are insensitive to $\sz$ errors of the form $\epsilon\h_\mathrm{noise} = \epsilon_z \sz$. 

Assuming that the gates are implemented by rectangular pulses, $\h_\mathrm{ideal}(t) = \frac{\pi}{\tau} \sx$ for $0\leq t <\tau$ and  $\h_\mathrm{ideal}(t) = -\frac{\pi}{\tau} \sx$ for $\tau\leq t <2\tau$,  where $\tau$ is the gate duration. Including the $\epsilon_z \sz$ noise contribution, the total Hamiltonians for $0\leq t <\tau$ and $\tau\leq t <2\tau$ are given by
\begin{equation}
\begin{split}
\h_+ &= \frac{\pi}{\tau} \{  \sx + \epsilon_z \sz\},\\
\h_- &= \frac{\pi}{\tau} \{ - \sx + \epsilon_z \sz\}\\
\end{split}
\end{equation}
respectively. The unitary operator corresponding to this pseudoidentity can be written as 
\begin{equation}
U = e^{-i\h_- \tau}e^{-i\h_+ \tau}  = \left(
\begin{array}{cc}
 1-i\pi\epsilon^3 + \mathcal{O}(\epsilon^5)& \mathcal{O}(\epsilon^5) \\
 \mathcal{O}(\epsilon^5) &1+i\pi\epsilon^3 + \mathcal{O}(\epsilon^5)
  \end{array}
  \right).
\end{equation}
The lowest degree of $\epsilon$ that appear is the above equation is $3$, showing the insensitivity of the pseudoidentity operation to $\sz$ errors. By symmetry, this also applies to any error of the form $\epsilon\h_\mathrm{noise} = \epsilon_y \sy + \epsilon_z \sz$ since we can perform a change of basis by rotating about the $\xax$ axis. The pseudoidentity itself is not insensitive to errors of the form $\epsilon\h_\mathrm{noise} = \epsilon_x \sx$, however. In the presence of a Hamiltonian error of the form $\epsilon_x \sx$, the noise Hamiltonian commutes with the driving Hamiltonian and the resulting pseudoidentity is trivially an $X(4\tau\epsilon_x)$ rotation. However, since the initial state of the qubit is along the $\hat{\textbf{x}}$ axis, $X$ rotations don't affect the qubit state because $\hat{X}(4\tau\epsilon_x)\ket{+} = \ket{+}$.

Since the dynamics are insensitive to error of the form $\epsilon\h_\mathrm{noise} = \epsilon_y \sy + \epsilon_z \sz$ and $\epsilon\h_\mathrm{noise} = \epsilon_x \sx$, they are in general insensitive to errors of the form $\epsilon\h_\mathrm{noise} = \epsilon_x \sx + \epsilon_y \sy + \epsilon_z \sz$. 

\section{Dynamics of an undriven qubit with the qubit-TLS noise model} \label{app:undriven_qubit_tls_solution}

Below we provide a solution to the Master equation of the undriven qubit in the presence of a $\nu_{zx} \sz\otimes \hat{\tau}_x$ interaction with a TLS. The total Hamiltonian, introduced in  \eqref{eq:qubit_tls_model_hamiltonian_equation}, has the form
\begin{equation}
    \hat{\mathcal{H}} = \delta \omega \sz + \nu_{zx} \sz \hat{\tau}_x, 
\end{equation}
where $\sz$ and $\hat{\tau}_x$ are the $z$ and $x$ Pauli operators for the qubit and TLS, respectively, and the kronecker product ``$\otimes$" is implied. We don't include amplitude damping noise on the qubit for increased clarity of the resulting equations, and we include the effects of qubit dephasing and amplitude damping on the TLS later in the derivation.

The evolution of the total density operator, neglecting qubit dephasing and TLS dissipation, is given by the Liouville von-Neumann equation
\begin{equation}
    \dv{}{t}\hat{\rho}(t) = -i[\hat{\mathcal{H}}, \hat{\rho}(t)].
\end{equation}
Solving this equation and noting that the qubit and qubit-TLS interaction terms commute, i.e. $[ \delta \omega \sz\otimes I , \nu_{zx} \sz \hat{\tau}_x] = 0$, gives 
\begin{equation}
    \hat{\rho}(t) = e^{-i\delta\omega \sz t} e^{-i\nu_{zx} \sz \hat{\tau}_x t} \bigl[ \hat{\rho}(0) \bigr]e^{i\nu_{zx} \sz \hat{\tau}_x t}e^{i\delta\omega \sz t}.
\end{equation}
Let the superscripts $S,B$ denote the subspaces corresponding to the qubit and the TLS respectively. We may obtain the qubit subsystem density matrix $\hat{\rho}^S$ by evaluating the partial trace over the TLS Hilbert space. 
\begin{equation}
    \hat{\rho}^S(t) = \mathrm{Tr}_B\left[\hat{\rho}(t)\right] = e^{-i\delta\omega \sz t} \mathrm{Tr}_B\left[e^{-i\nu_{zx} \sz \hat{\tau}_x t} \hat{\rho}(0)e^{i\nu_{zx} \sz \hat{\tau}_x t}\right] e^{i\delta\omega \sz^B t} 
\end{equation}
We now assume that the system and bath are initially unentangled, i.e. $\hat{\rho}(0) = \hat{\rho}^S(0)\otimes\hat{\rho}^B(0)$ and expand the partial trace over the TLS degree of freedom in terms of the eigenstates of the $\hat{\tau}_x$ operator, $\ket{\pm} = \frac{1}{\sqrt{2}}\left(\ket{0}\pm\ket{1}\right)$, with eigenvalues $\lambda_\pm=\pm 1$, respectively, giving
\begin{equation}
\begin{split}
    \hat{\rho}^S(t)
    &=  \sum_{\alpha\in \{+,-\}} \bra{\alpha}\hat{\rho}^B(0)\ket{\alpha} e^{-i(\delta\omega+\lambda_\alpha\nu_{zx})  \sz t} \hat{\rho}^S(0)e^{i(\delta\omega+\lambda_\alpha\nu_{zx}) \sz t}.
\end{split}
\end{equation}
We may obtain a matrix representation of the system density operator by expanding in terms of the eigenstates of the $\sz$ operator, giving
\begin{equation}
\begin{split}
    \hat{\rho}^S(t)&= \hat{\rho}^S_{00}(0) \ket{0}\bra{0} + \hat{\rho}^S_{11}(0) \ket{1}\bra{1} + f(t) \hat{\rho}^S_{01}(0) \ket{0}\bra{1} + f^*(t) \hat{\rho}^S_{10}(0) \ket{1}\bra{0} \\
    &= \left(\begin{array}{cc} \hat{\rho}^S_{00}(0) & f(t) \hat{\rho}^S_{01}(0) \\ f^*(t) \hat{\rho}^S_{10}(0) & \hat{\rho}^S_{11}(0) \end{array}\right)
\end{split}
\label{eq:qubit_tls_solution_intermediate_ft}
\end{equation}
where $f(t) = e^{-2i\delta\omega t} \left(\hat{\rho}^B_+(0) e^{-2i\nu_{zx}t}+\hat{\rho}^B_-(0) e^{2i\nu_{zx}t}\right)$, and where  $\hat{\rho}^S_{ij}(0) = \bra{i}\hat{\rho}^S(0)\ket{j}$ and $\hat{\rho}^B_\pm(0) = \bra{\pm}\hat{\rho}^B(0)\ket{\pm}$. As described in \secref{sssec:Qubit-TLSmodel}, we consider the case where the initial state of the TLS and qubit are $\hat{\rho}^B(0) = \ket{0}\bra{0}$ and $\hat{\rho}^S(0) = \ket{+}\bra{+}$ respectively. In this case, we obtain
\begin{equation}
\begin{split}
    \hat{\rho}^S(t)&= \frac{1}{2}\left(\begin{array}{cc} 1 & e^{-2i\delta\omega t} \cos(2\nu_{zx}t) \\ e^{2i\delta\omega t} \cos(2\nu_{zx}t) & 1 \end{array}\right).
\end{split}
\end{equation}

The equation above doesn't take qubit dephasing into account. The effect of dephasing noise acting on the qubit commutes with the effect of the Hamiltonian. Thus, since dephasing noise adds an exponential damping to the off-diagonal terms of $\hat{\rho}_S$ \cite{nielsen2002quantum}, the system density matrix in the presence of dephasing noise is given by
\begin{equation}
\begin{split}
    \hat{\rho}^S(t)&= \frac{1}{2}\left(\begin{array}{cc} 1 & e^{(-2i\delta\omega - 2\Gamma_D) t} \cos(2\nu_{zx}t) \\ e^{(2i\delta\omega -\Gamma_D)t} \cos(2\nu_{zx}t) & 1 \end{array}\right).
\end{split}
\label{eq:zx_interaction_solution}
\end{equation}

So far, we have not included amplitude damping on the TLS. In Ref. \cite{pangAbruptTransitionsMarkovian2017}, an equation describing the dynamics of a qubit coupled to a dissipative TLS was derived. Using those results, we find that the off-diagonal component $\hat{\rho}^S_{01}(t)$ in the presence of amplitude damping with strength $\kappa$ on the TLS is given by
\begin{equation}
\hat{\rho}^S_{01}(t) = \frac{1}{4}e^{(-2i\delta \omega - 2 \Gamma_D - \frac{\kappa}{4})t} \Bigl[ (e^{t\sqrt{\frac{\kappa^2}{16} - 4\nu_{zx}^2}  } + e^{-t\sqrt{\frac{\kappa^2}{16} - 4\nu_{zx}^2}}) + \frac{\kappa}{4}\frac{(e^{t\sqrt{\frac{\kappa^2}{16} - 4\nu_{zx}^2}} - e^{-t\sqrt{\frac{\kappa^2}{16} - 4\nu_{zx}^2}})}{\sqrt{\frac{\kappa^2}{16} - 4\nu_{zx}^2}}  \Bigr].
\label{eq:extended_qubit_tls_rho01}
\end{equation}
The diagonal components are unaffected, i.e. $\hat{\rho}^S_{00}(t) = \hat{\rho}^S_{00}(t) = \frac{1}{2}$. In Ref. \cite{pangAbruptTransitionsMarkovian2017}, such an equation was used to show that there is an abrupt transition from the system showing non-Markovian dynamics when $\frac{\kappa^2}{16} < 4\nu_{zx}^2$, corresponding to a low dissipation on the TLS, to the system showing purely Markovian dynamics when $\frac{\kappa^2}{16} > 4\nu_{zx}^2$, corresponding to a high dissipation on the TLS.

\section{Solution of the post Markovian master equation for the undriven qubit}
\label{app:pmme_zx_interaction}
In this section, we solve the PMME for the undriven qubit. This has been done in Ref. \cite{zhangPredictingNonMarkovianSuperconducting2021}, however, we include the derivation for our specific noise model here for completeness, and also because we require the full equations in order to relate the model to the qubit-TLS model. As before, for clarity of the resulting equations, we neglect amplitude damping on the qubit.

We want to solve the PMME, given by
\begin{equation}
  \begin{split}
    \dv{}{t} \hat{\rho}(t) =& \lin _0 \hat{\rho}(t) + \lin_1 \int_0^t dt'e^{-bt'} \text{exp}[(\lin_0 + \lin_1)t'] \hat{\rho}(t-t'),
  \end{split}
\end{equation}
where $\lin_0 \hat{\rho} = -i[\delta\omega \sz, \hat{\rho}]  + \Gamma_D(\sz \hat{\rho}\sz - \hat{\rho})$, $\lin_1 \hat{\rho} = \gamma_z(\sz \hat{\rho}\sz - \hat{\rho})$. To solve this equation, we first represent $\hat{\rho}(t)$ in an orthonormal basis containing the eigenoperators of $\lin_0 + \lin_1$.

\begin{equation}
  \begin{split}
\hat{\rho}(t) =& \sum_i \mu_i(t)\hat{R}_i,\\
\hat{R}_0 =& \frac{1}{\sqrt{2}}\left(\begin{array}{cc} 1 & 0\\ 0 & 1\end{array}\right), \hat{R}_1 = \left(\begin{array}{cc} 0 & 1\\ 0 & 0\end{array}\right),\hat{R}_2 = \left(\begin{array}{cc} 0 & 0\\ 1 & 0\end{array}\right),\hat{R}_3 = \frac{1}{\sqrt{2}}\left(\begin{array}{cc} 1 & 0\\ 0 & -1\end{array}\right),\\
  \end{split}
\end{equation}
where the operators $\hat{R}_i$ are eigenoperators of $\lin_0$ with eigenvalues $\lambda^0_i = \{0,-2i\delta\omega - 2\Gamma_D, 2i\delta\omega  - 2\Gamma_D,0\}$ and of $\lin_1$ with eigenvalues $\lambda^1_i = \{0,-2\gamma_z, -2\gamma_z,0\}$, and obey $\Tr{\hat{R}_i^T \hat{R}_j} = \delta_{ij}$.  Substituting these in the PMME, we get
\begin{equation}
  \begin{split}
    \sum_i\dv{\mu_i(t)}{t} \hat{R}_i =&  \sum_i \mu_i(t)\lin _0 \hat{R}_i + \sum_i\lin_1 \int_0^t dt'e^{-bt'}\mu_i(t-t') \text{exp}[(\lin_0 + \lin_1)t'] \hat{R}_i.
  \end{split}
\end{equation}
Projecting this equation onto each of the eigenoperators $\hat{R}_i$ (by right multiplying by $\hat{R}_i^T$ and taking the trace) we get 4 uncoupled equations
\begin{equation}
  \dv{\mu_i(t)}{t} =  \mu_i(t)\lambda^0_i + \lambda^1_i \int_0^t dt'e^{-bt'}\mu_i(t-t') e^{\lambda_i t'},
\end{equation}
where $\lambda_i = \lambda^0_i + \lambda^1_i$.  
Taking the Laplace transform of these equations, we obtain
\begin{equation}
s \tilde{\mu}_i(s)- \mu_i(0) = \lambda^0_i\tilde{\mu}_i(s) + \lambda^1_i \frac{1}{s-\lambda_i + b}\tilde{\mu}_i(s).
\end{equation}
Rearranging terms gives
\begin{equation}
\tilde{\mu}_i(s) = \frac{\mu_i(0)}{s-\lambda^0_i - \frac{\lambda^1_i }{s-\lambda_i+b}}.\\
\end{equation}
Finally, we can take the inverse Laplace transform, denoted $\text{Lap}^{-1}$, to obtain
\begin{equation}
\mu_i(t) = \text{Lap}^{-1}\Biggl[\frac{1}{s-\lambda^0_i - \frac{\lambda^1_i }{s-\lambda_i+b}}\Biggr] \mu_i(0).
\label{eq:inv_laplace_transform_mu_i}
\end{equation}
To perform the inverse Laplace transform, we can use the residue theorem. To do this, we first need to write the function to invert as a rational fraction, i.e.,
\begin{equation}
  \begin{split}
    \frac{1}{s-\lambda^0_i - \frac{\lambda^1_i }{s-\lambda_i+b}} = & \frac{s-\lambda_i+b}{(s-\lambda^0_i)(s-\lambda_i+b) - \lambda^1_i }\equiv \frac{P_i(s)}{Q_i(s)},\\
  \end{split}
\end{equation}
where $P_i(s)$ and $Q_i(s)$ are polynomials in $s$. Using the residue theorem with \eqref{eq:inv_laplace_transform_mu_i}, we get
\begin{equation}
  \mu_i(t) = \Biggl[
    \frac{P_i(z_{i+})}{Q_i'(z_{i+})} e^{z_{i+}t} + \frac{P_i(z_{i-})}{Q_i'(z_{i-})} e^{z_{i-}t}
  \Biggr] \mu_i(0),
  \label{eq:pmme_after_residue_theorem}
\end{equation}
where $z_{i\pm}$ are roots of $Q(s) = 0$ which are given by
\begin{equation}
z_{i\pm} = \lambda^0_i - \frac{b-\lambda^1_i}{2} \pm \sqrt{\frac{(b-\lambda^1_i)^2}{4}+\lambda^1_i}.
\end{equation}
Here we have assumed $\lambda^1_i \neq 0$. In the special case of $\lambda^1_i = 0$, we simply have $\mu_i(t) = e^{\lambda^0_i t}\mu_i(0)$.
We now substitute $\lambda^0_i = \{0,-2i\delta\omega - 2\Gamma_D, 2i\delta\omega- 2\Gamma_D,0\}$ and $\lambda^1_i = \{0,-2\gamma_z , -2\gamma_z,0\}$ to obtain 
\begin{equation}
\begin{split}
z_{1\pm} &= -2i\delta\omega  - 2\Gamma_D - \frac{b+2\gamma_z}{2} \pm \sqrt{ \frac{(b+2\gamma_z)^2}{4} - 2\gamma_z}\\
z_{2\pm} &= +2i\delta\omega - 2\Gamma_D - \frac{b+2\gamma_z}{2} \pm \sqrt{ \frac{(b+2\gamma_z)^2}{4} - 2\gamma_z}.\\
\end{split}
\end{equation}
The coefficients $\mu_0(t)$ and $\mu_3(t)$ are constant with time, i.e. $\mu_0(t) = \mu_0(0)$ and $\mu_3(t) = \mu_3(0)$.
For the qubit initialised in the state $\ket{+}$, $\mu_0(0) = \frac{1}{\sqrt{2}}$, $\mu_1(0) = \mu_2(0) = \frac{1}{2}$, $\mu_3(0) = 0$. Let $\gamma' = \frac{b+2\gamma_z}{2}$. We obtain 
\begin{equation}
\begin{split}
\mu_1(t) =& \frac{1}{4}e^{(-2i\delta \omega - 2 \Gamma_D - \gamma')t} \Bigl[ (e^{t\sqrt{\gamma'^2 -2\gamma_z}} + e^{-t\sqrt{\gamma'^2 -2\gamma_z}}) + \gamma'\frac{(e^{t\sqrt{\gamma'^2 -2\gamma_z}} - e^{-t\sqrt{\gamma'^2 -2\gamma_z}})}{\sqrt{\gamma'^2-2\gamma_z}}  \Bigr].
\label{eq:general_undriven_pmme_mu}
\end{split}
\end{equation}
In \eqref{eq:extended_qubit_tls_rho01}, we presented the equation for the off-diagonal component of the qubit subspace in the qubit-TLS model. Since $\mu_1(t) = \hat{\rho}^S_{01}(t)$, we have
\begin{equation}
\mu_1(t) = \frac{1}{4}e^{(-2i\delta \omega - 2 \Gamma_D - \frac{\kappa}{4})t} \Bigl[ (e^{t\sqrt{\frac{\kappa^2}{16} - 4\nu_{zx}^2}  } + e^{-t\sqrt{\frac{\kappa^2}{16} - 4\nu_{zx}^2}}) + \frac{\kappa}{4}\frac{(e^{t\sqrt{\frac{\kappa^2}{16} - 4\nu_{zx}^2}} - e^{-t\sqrt{\frac{\kappa^2}{16} - 4\nu_{zx}^2}})}{\sqrt{\frac{\kappa^2}{16} - 4\nu_{zx}^2}}  \Bigr].
\label{eq:extended_qubit_tls_mu}
\end{equation}
Comparing \ref{eq:general_undriven_pmme_mu} and \ref{eq:extended_qubit_tls_mu}, we can see that the two models can be related to each other via the relations
\begin{equation}
\begin{split}
\gamma' = &  \frac{\kappa}{4},\\
\gamma_z= &  2\nu_{zx}^2, \\
b = & \frac{\kappa}{2} - 4\nu_{zx}^2.
\label{eq:qubit_tls_pmme_parameter_relations}
\end{split}
\end{equation}
We analyse and interpret these relations in \secref{subsec:PMME_qtls_relation}.

\section{Over-parameterisation in the PMME and qubit-TLS model}
\label{app:overparameterisation}

In \secref{subsec:undriven_qubit_results} we show that it is not necessary to include dissipation on the TLS in order to fit the experimental results. In this section, we further discuss the over-paramterisation due to the excess parameter in the qubit-TLS and the PMME model.

\begin{figure}[ht]
  \includegraphics[width=246pt]{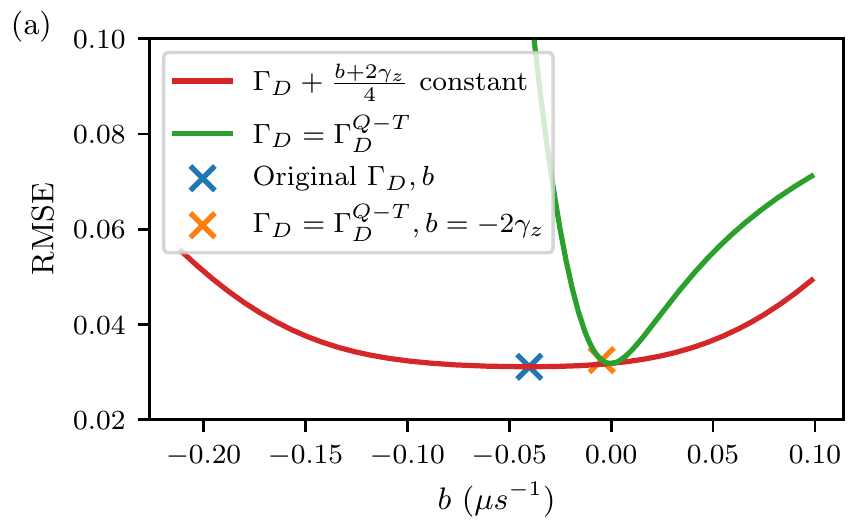}
  \includegraphics[width=246pt]{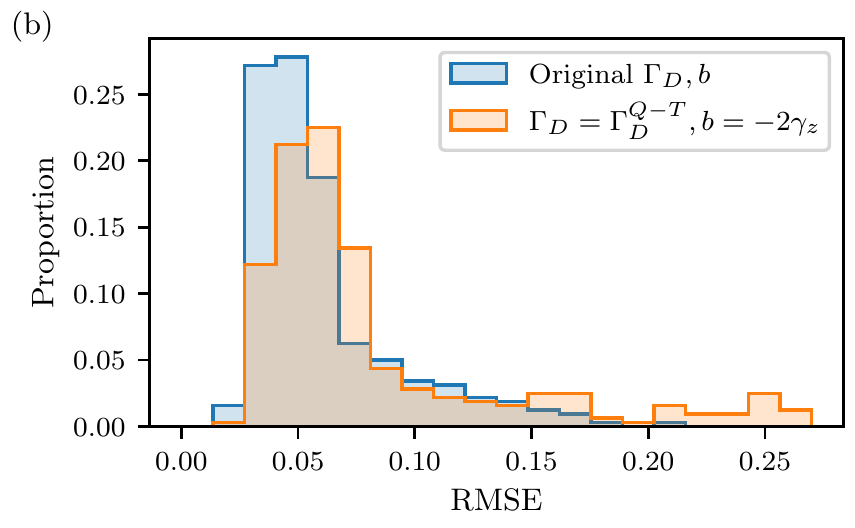}
  \caption{
  (a) The root-mean-squared errors (RMSE) plotted against $b$ for a single experiment. The red curve shows the RMSE when $b$ and $\Gamma_D$ are both changed while keeping $\Gamma_{D} + \frac{b+2\gamma_z}{4}$ constant, while the red curve shows the RMSE when $\Gamma_D = \Gamma_D^{Q-T}$. The blue cross denotes the RMSE of the original, optimal fitting parameters and the orange cross denotes the RMSE when $b$ and $\Gamma_D$ are set to specific values. The plateau in the red curve suggests over-parameterisation of the model due to insensitivity of the fitting quality. 
  (b) Distribution of the RMSE of the PMME model for the undriven qubit when the noise model parameters obtained from non-linear regression are left unchanged (blue) and when they are manually set (orange). In the latter case, $\Gamma_D$ is set to the corresponding value in the qubit-TLS model, i.e. $\Gamma_D^{Q-T}$, while $b$ is set to $-2\gamma_z$. The values of $\gamma_z, \delta\omega, \Gamma_{AD}$ are left unchanged. We can see that hand-picking the noise model parameter values does not significantly affect the quality of the obtained results.}
  \label{fig:overparameterisation_plots}
\end{figure}

The PMME model has the same number of parameters as the qubit-TLS model. Thus, an excess parameter in the qubit-TLS model implies that the PMME model is over-parameterised as well.
Using \eqref{eq:general_undriven_pmme_mu}, we find that in the limit $\gamma'^2 \ll |\gamma_z|$, which corresponds to weak dissipation on the TLS, the PMME model dynamics are given by
\begin{equation}
\begin{split}
\mu_1(t) =& \frac{1}{4}e^{(-2i\delta \omega - 2 \Gamma_D - \frac{b}{2} - \gamma_z)t} \Bigl[ e^{it\sqrt{2\gamma_z}} + e^{-it\sqrt{ 2\gamma_z}}   \Bigr],
\end{split}
\end{equation}
and the qubit-TLS dynamics are given by
\begin{equation}
\mu_1(t) = \frac{1}{4}e^{(-2i\delta \omega - 2 \Gamma_D - \frac{\kappa}{4})t} \Bigl[ (e^{it\sqrt{4\nu_{zx}^2}  } + e^{-it\sqrt{4\nu_{zx}^2}})\Bigr].
\end{equation}

Comparing the above equations, we can see that although $\gamma_z$ is set to be $2\nu_{zx}^2$, there is a freedom in the values of $\Gamma_D$ and $b$ because various choices of those parameters will satisfy $2\Gamma_D + \frac{b}{2} +\gamma_z= 2\Gamma_D^\mathrm{Q-TLS} + \frac{\kappa}{4}$, where $\Gamma_D^\mathrm{Q-TLS}$ is the dephasing rate in the qubit-TLS model. Thus, the PMME model is over-parameterised in this case.
To see this over-parameterisation, in \figref{fig:overparameterisation_plots} (a) we show a representative example of how the fitting quality varies with the value of $b$. When $b$ and $\Gamma_D$ are changed while keeping the effective dephasing rate $\Gamma_{D} + \frac{b+2\gamma_z}{4}$ constant, the RMSE has a plateau and the fitting quality is insensitive to the value of $b$. However, if we change $b$ while fixing $\Gamma_D$ to be equal to $\Gamma_D^{Q-T}$, we observe a sharp minimum in RMSE. This shows that the noise model is over-parameterised in this region.

Applying the constraint $b = -2\gamma_z$, which corresponds to $\kappa = 0$, results in a model that does not suffer from this over-parameterisation. In order to analyse how the obtained results are affected when this constraint is applied, instead of performing the non-linear regression again with this constraint, we manually change the parameters obtained from the existing regression results. We set the the value of $\Gamma_D$ in the PMME model to the value of $\Gamma_D$ in the qubit-TLS model. We set the value of $b$ to be equal to $-2\gamma_z$ where $b,\gamma_z$ are both the PMME model parameters. The values of $\gamma_z, \delta\omega, \Gamma_{AD}$ in the PMME model are left unchanged. In \figref{fig:overparameterisation_plots} (a), the blue and orange crosses correspond to the unconstrained and the constrained results respectively and we can observe that the corresponding values of the RMSE are very similar.

In \figref{fig:overparameterisation_plots} (b), we compare the distribution of the root-mean-squared-errors for the original PMME results and the results with the manually applied constraint for all the undriven qubit results. We observe that the quality of solutions does not significantly worsen when we change the model parameters as described above. The median RMSE increases from $\sim 0.05$ to $\sim 0.06$. Repeating the non-linear regression with the constraint applied can only lead to improved results compared to results that we have shown here in which apply the constraint manually.

\section{Relating the PMME with the qubit-TLS model}
\label{app:pmme_qtls_relation}

\begin{figure}
  \includegraphics[width=0.5\columnwidth]{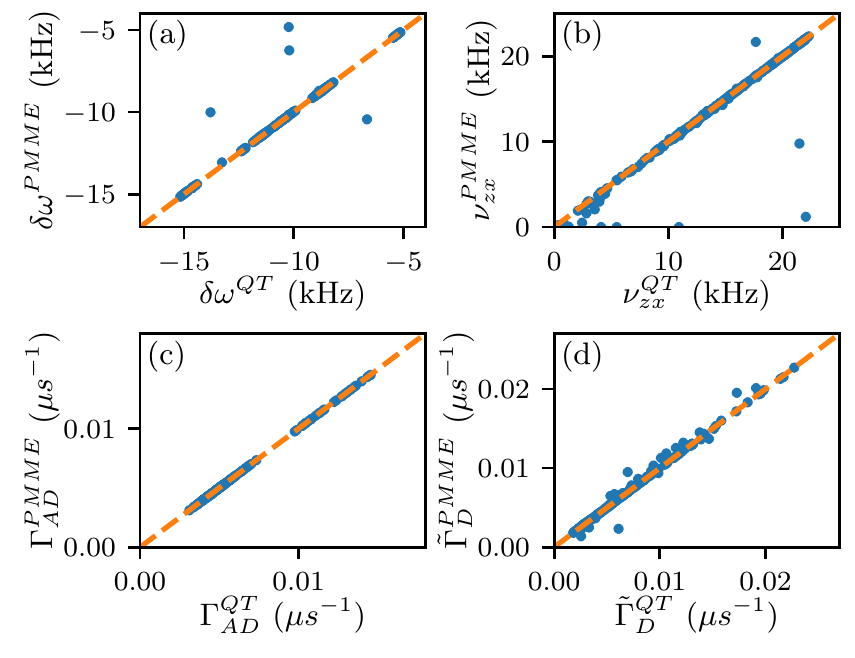}
  \caption{Comparison of the effective PMME model parameters (vertical axes) with the noise model parameters obtained for the qubit-TLS model (horizontal axes). The orange dashed line has slope $1$, starts at the origin, and has been added to guide the eye.}
  \label{fig:pmme_vs_qtls_id}
\end{figure}
In this section, we numerically confirm the validity of the relations in \eqref{eq:qubit_tls_pmme_parameter_relations} in \figref{fig:pmme_vs_qtls_id}, where we compare the noise model parameters obtained for the qubit-TLS model with the corresponding effective parameters of the PMME model discussed in \secref{subsec:PMME_qtls_relation}. Due to the over-parameterisation discussed in \appref{app:overparameterisation}, instead of comparing the dephasing model parameters, we compare the effective dephsing rate given by 
\begin{equation}
\tilde{\Gamma}_D^{PMME} = \Gamma_D + \frac{b+ 2\gamma_z}{4}
\end{equation}
for the PMME model and 
\begin{equation}
\tilde{\Gamma}_D^{QT} = \Gamma_D^\mathrm{Q-TLS} + \frac{\kappa}{8}
\end{equation}
for the qubit-TLS model. 
We can see that the parameters do indeed follow the linear relationship predicted by \eqref{eq:qubit_tls_pmme_parameter_relations}. This shows that the non-linear regression is robust, since the sets of parameters for qubit-TLS and PMME are obtained independently.

\section{Markovian Master Equation Simulation for repeated pseudoidentities}
\label{app:markovian_me_simulation}
We describe here our method for the efficient noisy simulation of quantum circuits with repeated pseudoidentities.
We represent the density matrix in the Pauli basis such that $\hat{\rho} = (\hat{I} + c_x\sx+ c_y\sy+ c_z\sz)/2$. Let $\boldsymbol{c} = \{1,c_x,c_y,c_z\}$. Denote the state before the initial state preparation as $\boldsymbol{c}_\mathrm{initial}$ and the final state after the measurement basis change as $\boldsymbol{c}_\mathrm{final}$. We have
\begin{equation}
\begin{split}
\boldsymbol{c}_\mathrm{final} =&  \Lambda_\mathrm{basis-change} \Lambda_\mathrm{pseudoidentity}^n\Lambda_\mathrm{state-prep} \boldsymbol{c}_\mathrm{initial}, 
\end{split}
\end{equation}
where $\Lambda_\alpha$ denote the corresponding superoperators. Each of the superoperators can be the result of multiple quantum gates. Since we assume that the Hamiltonian corresponding to each quantum gate is constant, the Hamiltonian $\h(t)$ is piece-wise constant. Since the Hamiltonian and Lindblad jump operators are constant during each gate, the superoperator corresponding to a gate sequence of $m$ gates can be written as:
\begin{equation}
\Lambda = \Lambda_m \Lambda_{m-1} ... \Lambda_1 = e^{\lin_m \Delta t_m} e^{\lin_{m-1} \Delta t_{m-1}} ... e^{\lin_1 \Delta t_1},
\end{equation}
where $\lin_i, \Delta_i$ are the time-independent Lindbladian and duration of gate $i$, respectively. We now describe how we find the matrix representation of the Lindbladian $\lin_i$. Let $\hat{F}_i$ denote the basis $\{\hat{I},\sx,\sy,\sz\}$ which has the property $\Tr[\hat{F}_j\hat{F}_i] = 2\delta_{ij}$. We first expand the Lindblad Master equation in this basis to obtain:
\begin{equation}
\begin{split}
     \sum_l \dot{c_l} \hat{\sigma}_l = -i\sum_i c_i [\h,\hat{\sigma}_i] + \sum_i c_i \sum_k \Gamma_k \qty(\lhat_k\hat{\sigma}_i \lhat_k^\dagger - \frac{1}{2}\qty{\lhat_k^\dagger \lhat_k, \hat{\sigma}_i}).\\
\end{split}
\end{equation}
After left-multiplying by $\hat{F}_j$ and taking the trace, we obtain
\begin{equation}
\begin{split}
     \sum_l \dot{c_l} \Tr[\hat{\sigma}_j \hat{\sigma}_l] &= -i\sum_i c_i \Tr[\hat{\sigma}_j [\h,\hat{\sigma}_i]] + \sum_i c_i \Tr[\sum_k \Gamma_k \hat{\sigma}_j \qty(\lhat_k\hat{\sigma}_i \lhat_k^\dagger - \frac{1}{2}\qty{\lhat_k^\dagger \lhat_k, \hat{\sigma}_i})].\\
\end{split}
\end{equation}
Using $\Tr[\hat{F}_j\hat{F}_i] = 2\delta_{ij}$, this simplifies to 
\begin{equation}
\begin{split}
    \ \dot{c_j}  &= -i\sum_i c_i h_{ji} + \sum_i c_i D_{ji}.\\
\end{split}
\end{equation}
In the vector representation, we have
\begin{equation}
\begin{split}
     \dot{\boldsymbol{c}}  &= (-i\boldsymbol{h}  + \boldsymbol{D})\boldsymbol{c}.\\
\end{split}
\end{equation}
The solution to this is given by
\begin{equation}
\begin{split}
    \boldsymbol{c}(t)  &= e^{(-i\boldsymbol{h}  + \boldsymbol{D})t}\boldsymbol{c}(0) = e^{\boldsymbol{\ell}t}\boldsymbol{c}(0), \\
\end{split}
\end{equation}
where $\boldsymbol{\ell}$ is the matrix representation of the Lindbladian $\lin$.

After finding the superoperator corresponding to the initial state preparation, single pseudoidentity, and the measurement basis change, we diagonalise the superoperator corresponding to a single pseudoidentity such that $\Lambda_{\text{pseudoidentity}} = UVU^{-1}$, where $U$ and $V$ are the matrix of eigenvectors and eigenvalues of $\Lambda_{\text{pseudoidentity}}$, respectively. Thus, for any integer $n$, $\Lambda_{\text{pseudoidentity}}^n = UV^nU^{-1}$ and the final state of the qubit $\boldsymbol{c}_\mathrm{final}$ can be efficiently obtained.

\section{Non-linear regression uncertainty estimation}
\label{app:uncertainty_estimation}
To obtain uncertainties in the estimates of the ratio of error parameters, we first estimate uncertainties in the fitting parameters as follows
\begin{enumerate}    
  \item Using finite difference, calculate the Jacobian matrix $J$ of the loss function $\mathscr{L}$ at the obtained minima;
  \item Estimate the covariance matrix from $J$ using $C  = (J^T J)^{-1} \mathscr{L}_\mathrm{min}$ \cite{wright1999numerical};
  \item Obtain the uncertainties $\sigma_i$ using the covariance matrix via $\sigma_i = \sqrt{C_{ii}}$.
\end{enumerate}
After finding the uncertainty associated with each fitting parameter, we can use textbook propagation of error techniques to find the uncertainty associated with the ratio of fitting parameters, e.g. $\frac{\delta \omega (\thetagate)}{\delta \omega (0)}$.